\documentclass[aps,prd,amsmath,amssymb,10pt,letterpaper,balancelastpage,showpacs,superscriptaddress,nofootinbib,floatfix,notitlepage,preprintnumbers,longbibliography,onecolumn]{revtex4-2}

\usepackage{graphicx}
\usepackage{soul}
\usepackage{color}
\usepackage{xcolor}
\usepackage{enumitem}
\usepackage{cancel}
\usepackage{mathtools}
\usepackage{hyperref}
\usepackage{orcidlink}
\usepackage{cases}
\usepackage[normalem]{ulem}
\usepackage{bbm}

\usepackage[T1]{fontenc}

\usepackage{tabularray}
\usepackage{multirow}

\usepackage[export]{adjustbox}

\usepackage{amsthm}

\theoremstyle{remark}

\theoremstyle{remark}

\theoremstyle{plain}

\usepackage{tabularray}
\UseTblrLibrary{booktabs}

\newcommand{\pfrac}[2]{{\left(\frac{#1}{#2}\right)}}

\newcommand{\Var}{{\rm Var}}

\newcommand{\bra}[1]{\left\langle #1 \right|}
\newcommand{\ket}[1]{\left|#1\right\rangle}
\newcommand{\braket}[2]{\left\langle#1 \middle|  #2\right\rangle} 
\newcommand{\braopket}[3]{\left\langle #1 \middle| #2 \middle| #3 \right\rangle} 
\newcommand{\expect}[1]{\left\langle#1\right\rangle}

\newcommand{\Tr}{{\rm Tr}}

\newcommand{\Hbos}{\mathcal{H}_{\rm boson}}
\newcommand{\Hqub}{\mathcal{H}_{\rm qubit}^{(N)}}
\newcommand{\Hsym}{\mathcal{H}_{\rm sym}^{(N)}}

\newcommand{\ketF}[1]{\left|#1\right\rangle_{\rm B}}
\newcommand{\ketQ}[1]{\left|#1\right\rangle_{\rm Q}}

\newcommand{\ketD}[1]{\left|#1\right\rangle_{\rm Q}}
\newcommand{\ketB}[1]{\left|#1\right\rangle_{\rm Q}} 
\newcommand{\ketC}[1]{\left|#1\right\rangle_{\rm B}}

\newcommand{\GHZxN}[1]{\ket{\rm{GHZ}^x_{#1}}}
\newcommand{\catN}[1]{\ket{\rm{cat}_{#1}}}

\newcommand{\HPT}{\xleftrightarrow[]{{\rm~~HPT~~}}}
\newcommand{\HPTnp}{\xleftrightarrow[{\rm S~pole}]{{\rm~~HPT~~}}} 
\newcommand{\HPTeq}{\xleftrightarrow[{\rm equator}]{{\rm~~HPT~~}}}
\newcommand{\HPTapp}{\xleftrightarrow[{\rm approx.}]{{\rm~~HPT~~}}}

\begin{document}

\preprint{FERMILAB-PUB-26-0781-SQMS-T}

\title{A Correspondence Between Qubit- and Boson-Based Sensing Protocols}
\date{\today}

\author{Zachary Bogorad\,\orcidlink{0000-0001-9913-6474}\,}
\email{zbogorad@fnal.gov}
\affiliation{Fermi National Accelerator Laboratory, Batavia, IL 60510, USA}

\author{Roni Harnik\,\orcidlink{0000-0001-7293-7175}\,}
\email{roni@fnal.gov}
\affiliation{Fermi National Accelerator Laboratory, Batavia, IL 60510, USA}

\begin{abstract}
    \noindent Recent years have seen a surge of interest in adapting non-classical states to search for physics beyond the Standard Model, including both qubit- and boson-based systems. These two classes of systems are closely related through the Holstein--Primakoff transformation (HPT). In this work, we show that the HPT offers novel insights into inherently quantum-mechanical detection of new physics. We begin by discussing how the HPT can be used to unify certain existing qubit- and boson-based proposals in a single framework. We then show how it naturally reveals a novel generalization of GHZ states which offer superior performance for the detection of rare interactions. We demonstrate that these novel states can be readily prepared, and combine the coherent advantage of GHZ states with superior error rejection. Finally, we show that the HPT also reveals a natural extension of the Hamiltonian-Not-in-Lindblad-Span theorem for collections of bosonic modes.
\end{abstract}

\maketitle

\section{Introduction}

The search for wave-like dark matter is a well-motivated opportunity for quantum sensors. Modern quantum devices are able to control and measure the quantum state of diverse physical systems. These quantum states are often comprised of photons or elements that couple to photons, and can thus be used to search for dark matter candidates that interact with light, such as dark photons~\cite{Holdom1986} and axions~\cite{PecceiQuinn1977}. Classical resonant electromagnetic devices that are well-suited for this task include microwave cavities~\cite{Sikivie1983}, LC circuits~\cite{Sikivie1985}, and spin ensembles~\cite{Budker2014}. Quantum realizations of these systems can enhance the sensitivity of dark matter searches. Such quantum sensing protocols can be realized both within single modes of indistinguishable bosons, and within ensembles of distinguishable, finite-dimensional systems such as qubits. In this work, we draw connections between dark matter searches in these two categories. 

In bosonic systems, as an example, squeezed light has been used to enhance the scan speed in axion haloscopes~\cite{Malnou2021}. Photon counting in quantum controlled superconducting radio-frequency (SRF) cavities has been used to surpass the standard quantum limit in dark photon searches~\cite{Dixit:2020ymh}. In addition, states with non-zero bosonic occupation numbers can be used to stimulate the signal rates, either with Fock states~\cite{Agrawal:2023umy} or cat states~\cite{Zheng:2025qgv}. These protocols can be extended and generalized to utilize $m$ distingushable bosonic modes, say $m$ SRF cavities~\cite{Freiman:2025tse}. On the qubit side, dark matter search protocols include searching for excitations in unentangled qubits~\cite{Chen:2022quj, Chen:2024aya, Fukuda:2025afi}, searching for coherent excitation of qubit ensembles~\cite{Chen:2025tgj}, using spin coherent states~\cite{SuperradiantDetection1, Kitano:2026dgx}, as well as using maximally entangled states such as GHZ states~\cite{Chen:2023swh, Ito:2023zhp, Bodas:2025vff}; see Ref.~\cite{Moroi:2026ges} for a review. Within this zoo of protocols, one can achieve various scalings of the signal (and noise) rates with the figures of merit of the system. For example, the signal rate in the protocols of Refs.~\cite{Agrawal:2023umy,Zheng:2025qgv} scale as $mnV$, for a system of $m$ cavities, each with a volume $V$, and a total occupation number $n$. The signal rates in protocols that utilize maximally entangled qubits, such as Refs.~\cite{Chen:2023swh, Ito:2023zhp, Bodas:2025vff}, scale as $N^2$ for $N$ qubits.

In this work we will utilize a known correspondence between bosonic and distinguishable-spin systems---the Holstein-Primakoff transformation~\cite{HolsteinPrimakoff1940} (HPT), which was identified in a very different context---to understand the scaling behaviors of various sensing protocols. In particular, we will show that the HPT maps operators that act in the Hilbert space of a bosonic mode to ones that operate in the symmetric Hilbert subspace of distinguishable qubits. Once we define a fiducial ``vacuum'' state in both Hilbert spaces, the map between operators can then be used to produce identifications between states, as we will show in several examples. For example, bosonic Fock states are HPT counterparts of Dicke qubit states, and the same can be said for bosonic cat states and qubit GHZ states. Furthermore, we will show that the HPT can be used to draw correspondences between sensing protocols.

Applying this understanding to the landscape of DM sensing is beneficial in several ways. First, we will develop a more unified understanding of the connections between the scaling of different protocols, where the optimal quantum Fisher information can quite generically be written as
\begin{equation}
    F_Q^{\rm opt} \sim F_Q^{(0)} \times \left(\mbox{number of sensors}\right)
    \times \left(\mbox{total occupation number}+1\right)
\end{equation}
where $F_Q^{(0)}$ is the quantum Fisher information one would get with a single, unoccupied sensing component. This formula says that the sensitivity of a single empty sensor can be enhanced in two ways: increasing $N$ or $m$, the number of sensors, and increasing $n$, the occupation number in the whole system. In bosonic systems $m$ and $n$ can scale independently, giving a $\propto mn$ scaling. In systems of qubits (or qudits), where the Hilbert space per sensor is finite, the maximal effective occupation number also scales with $N$, giving an overall optimal scaling $\propto N^2$.

In fact, for a broad class of signals, our full scaling result will have an even stronger implication: unless decays are entirely negligible, bosonic detectors' sensitivities cannot be parametrically enhanced through the preparation of high-occupation number states; the only possible scaling is with mode number $m$, rather than with occupation $n$. This motivates the continued development of highly coherent devices with slower decay processes.

The second benefit  of the HPT understanding is that it will allow us to import protocols that have been developed for bosonic systems to equivalent searches on the distinguishable qubit side, and vice versa. In particular we will introduce a new qubit-based protocol based on a ``GHZ-4'' state, which retains the sensitivity of the standard GHZ state while allowing one to fully discriminate noise coming from qubit decays, a marked improvement compared to the GHZ (``GHZ-2'') protocols in Refs.~\cite{Chen:2023swh, Ito:2023zhp, Bodas:2025vff}. The GHZ-4 protocol is derived as the HPT partner of the 4-legged-cat protocol in bosonic systems~\cite{Zheng:2025qgv}. 

This work is organized as follows. In Section~\ref{sec:HPT} we review the Holstein-Primakoff transformation as a map between bosonic and (symmetric) qubit Hilbert spaces, both at the state and operator level. In Section~\ref{sec:examples} we show examples of pairs of states, on both sides of the HPT, including pictorial maps of phase state distributions to build intuition. In Section~\ref{sec:implications} we discuss the implications of the HPT for dark matter sensing protocols, including mappings between several known protocols. In Section~\ref{sec:DetectionUsingGHZ4} we present a new search protocol using the GHZ-4 state---the HPT analogue of a 4-legged-cat state---that allows for discrimination of qubit decay backgrounds. In Section~\ref{sec:NmScaling} we use the HPT to import scaling sensitivity bounds from qubit to bosonic systems, including the Hamiltonian-not-in-Lindblad-Span (HNLS)~\cite{Zhou2018HNLS} condition for achieving the Heisenberg Limit. We conclude in Section~\ref{sec:summary}.

\section{The Holstein--Primakoff Transformation}\label{sec:HPT}

\begin{center}
\begin{table}[b]
    \begin{tabular}{c @{\hspace{5pt}}|@{\hspace{5pt}} c}
    \multicolumn{2}{c}{\textbf{Examples of Qubit- and Boson-Based Sensing}} \\
    \hline
    \textbf{Qubit/Qudit Sensors} & \textbf{Bosonic Sensors} \\
    \hline
    Internal levels of ultracold atoms \cite{InterferometerCoherenceReview, Gross2010, Sewell2012, Colombo2022} & Electromagnetic cavities \cite{Raimond:2001zz, Blais:2020wjs, Dixit:2020ymh, Backes2021} \\
    Superconducting circuits (e.g. transmons) \cite{Krantz:2019jkw, Blais:2020wjs, Chen:2022quj, Chen:2024aya} & Quantum electron cyclotrons \cite{Brown:1985rh, Fan:2022uwu} \\
    Atom interferometers \cite{Cronin:2009zz, Bongs:2019exm, InterferometerCoherenceReview} & Levitated microspheres \cite{Gonzalez-Ballestero:2021gnu, Moore:2020awi} \\
    {} & Mechanical resonators \cite{Aspelmeyer:2013lha, Poot:2011hps, Carney:2020xol}
    \end{tabular}
    \caption{A variety of examples of quantum sensing platforms that are most naturally described as qubit/qudit-based or boson-based, at least for our purposes. Laser interferometers, which can reasonably be described in either way, are omitted here; see the text.}
    \label{tab:QubitBosonExamples}
\end{table}
\end{center}

Our focus throughout this work will be on connecting two classes of systems: those based on ``qubits'' (or, later in the work, ``qudits'')---by which we mean detection schemes using $N$ copies of a two-level (or, for qudits, $d$-level) system---and those based on ``bosons''---by which we mean a single bosonic mode (or, later in the work, $m$ parallel bosonic modes). This distinction is not always unambiguous however; we discuss an example of this below. The crucial distinguishing feature of the two system types will ultimately be the presence of a conserved number: the number $N$ of qubits will always remain unchanged throughout the observation process, whereas the total number of excitations $n$ in bosonic modes will vary.

We present a number of examples of qubit- and boson-based sensing platforms in Table~\ref{tab:QubitBosonExamples}. On the qubit side, pairs of internal levels of trapped ultracold atoms~\cite{InterferometerCoherenceReview, Gross2010, Sewell2012, Colombo2022}---hyperfine states, for example---form natural 2-level systems, while a variety of superconducting circuits can implement ``artificial atoms''~\cite{Krantz:2019jkw, Blais:2020wjs, Chen:2022quj, Chen:2024aya}. In both cases, there are typically infinitely-tall towers of states in principle, but these can be treated as qubits when the energy splitting to higher levels are sufficiently different (and thus sufficiently detuned from the signal) from them to be omitted. Atom interferometers~\cite{Cronin:2009zz, Bongs:2019exm, InterferometerCoherenceReview}, although they still use ultracold atoms, are a meaningfully different platform, as the distinct states are now branches of the interferometer (typically spatially separated), rather than internal states. Notably, neither atomic platform assumes that the atoms are fermionic: bosonic atoms will still have a conserved number, and thus still fall into the qubit/qudit category.

Electromagnetic cavities, including both normal- and superconducting cavities, are likely the most widespread bosonic sensing platforms~\cite{Raimond:2001zz, Blais:2020wjs, Dixit:2020ymh, Backes2021}; these typically employ only a single mode, or perhaps a few modes, at a time. A variety of other systems have internal bosonic excitations that can also be used for sensing, however, including the discretized energy levels of quantum electron cyclotrons~\cite{Brown:1985rh, Fan:2022uwu}, the motion of levitated microspheres~\cite{Gonzalez-Ballestero:2021gnu, Moore:2020awi}, and the phonons within a mechanical resonator, such as a cantilever~\cite{Aspelmeyer:2013lha, Poot:2011hps, Carney:2020xol}.

One widespread system can be described in either qubit or boson language, however: the laser interferometer~\cite{Danilishin:2012fa, Yurke:1986ipz, LIGOO4Detector:2023wmz, Mckenzie:2002ir}. It is natural to describe a laser interferometer as a bosonic system with $m=2$, modes, corresponding to the two arms of the interferometer. At the same time, a laser interferometer is mathematically quite similar to an atom interferometer, and can be described as a qubit system in much the same way, with each photon assigned to one of two arms/states; number conservation then holds so long as photon loss between the interferometer's beamsplitters (i.e. in the observation region) is negligible. The only subtlety in this case is that a laser interferometer will typically be in a superposition of many different photon counts, but since there is no interference between different photon numbers in the absence of loss, one can simply handle each photon number $n$ as its own $N$-qubit system and then average appropriately at the end. Either the qubit or boson description may be more convenient in different contexts.

The analogy between Fock states and Dicke states rests on a precise algebraic map between their corresponding Hilbert spaces, made explicit by the Holstein--Primakoff transformation (HPT)~\cite{HolsteinPrimakoff1940}. 
Holstein and Primakoff developed the map while studying the magnetization in ferromagnetic domains and its dependence on an applied external magnetic field. To that end the canonically normalized bosonic modes consisting of collective spin excitation were identified, providing a map between a (truncated) bosonic Hilbert space and the underlying spin modes. In our context we will take the map as a mathematical correspondence between two distinct sensing systems, one consisting of a bosonic mode and the other of distinguishable qubits which are symmetrized.
We begin by introducing the two spaces, then state the transformation and some limiting cases. Specific examples of the resulting correspondence are in the next section.

\subsection{Hilbert spaces}

Since a single bosonic mode can be fully described by its coefficients in the Fock basis, it occupies the Fock space
\begin{equation}
    \Hbos = \bigoplus_{k=0}^{\infty} \mathbb{C}\ketF{k} \cong \ell^2(\mathbb{N}_0),
\end{equation}
with $\ell^2(\mathbb{N}_0)$ the space of complex sequences $c_k$ for integer $k \geq 0$ whose norms have a finite sum, i.e. $\sum_k |c_k|^2 < \infty$; here the $c_k$ can be understood simply as the coefficients of the $k$'th Fock state $\ketF{k}$. Notably there is no limit on the occupation number of a single bosonic mode; this will be an important difference from the qubit case below.

The full Hilbert space of $N$ distinguishable qubits (i.e. spin-$1/2$ or two-level systems; we use all three terms interchangeably) is $\Hqub = (\mathbb{C}\ketQ{0} + \mathbb{C}\ketQ{1})^{\otimes N} = (\mathbb{C}^2)^{\otimes N}$. In this work, however, we are interested in the \emph{symmetric subspace}
\begin{equation}
    \Hsym = {\rm Sym}^N(\mathbb{C}^2)
    \subset \Hqub,
\end{equation}
which consists of all states invariant under permutation of the qubit labels. It is spanned by the $N+1$ Dicke states $\{\ketD{J,M}\}_{M=-J}^{J}$ with $J = N/2$, indexed by their eigenvalues $J(J+1)$ and $M$ under the operators for the total spin-squared $\mathbf{J}^2 = \mathbf{J}_x^2 + \mathbf{J}_y^2 + \mathbf{J}_z^2$ and for the total spin along a chosen (conventionally $z$) axis $\mathbf{J}_z$, respectively. (We will work in the convention that $\ketQ{0}$ is a spin down along $z$, with $\ketQ{1}$ spin up along $z$.) A state in $\Hsym$ carries no information about which qubits are in which state; only the total number of qubits in $\ketQ{1}$, i.e.\ $k = J + M \in \{0,\ldots,N\}$, is physical. This is the qubit analog of bosonic indistinguishability, but with the key difference noted above: $k$ is bounded from above by $N$.

\subsection{The Transformation}

The Holstein-Primakoff transformation is then an exact embedding of $\Hsym$ into $\Hbos$. It is defined by the following correspondence between qubit and boson operators:
\begin{subequations}
\begin{align}
    \mathbf{J}_z &\HPT   a^\dagger a -\frac{N}{2} =  \hat{n} - J \label{eq:HPTGeneral1} \\
    \mathbf{J}_+ &\HPT a^\dagger \sqrt{N - a^\dagger a} = a^\dagger \sqrt{N - \hat{n}} \label{eq:HPTGeneral2} \\
    \mathbf{J}_- &\HPT \sqrt{N - a^\dagger a} \, a = \sqrt{N - \hat{n}} \, a\label{eq:HPTGeneral3}
\end{align}
\end{subequations}
where $\mathbf{J}_\pm = \mathbf{J}_x \pm i\mathbf{J}_y$, $a$ ($a^\dagger$) is the lowering (raising) operator in $\Hbos$, and $\hat{n} = a^\dagger a$. It is straightforward to confirm that these bosonic operators then satisfy the usual $\mathfrak{su}(2)$ commutation relations. One can then identify the basis states via
\begin{align}
    \ketD{J,M} \longleftrightarrow \ketF{J+M}, \label{eq:DickeFockCorrespondence}
\end{align}
and in particular we have
\begin{subequations}
\begin{align}
    \ketQ{0}^{\otimes N} = \ketD{J,-J} &\longleftrightarrow \ketF{0} \\
    \ketQ{1}^{\otimes N} = \ketD{J,J} &\longleftrightarrow \ketF{N}.
\end{align}
\end{subequations}
Note that $\mathbf{J}_- \ketD{J,-J} = 0 = \sqrt{N - a^\dagger a} \, a \ketF{0}$ and likewise $\mathbf{J}_+ \ketD{J,J} = 0 = a^\dagger \sqrt{N - a^\dagger a} \ketF{N}$, so these re-weighted raising and lowering operators limit the bosonic space to the span of the Fock states from $\ketF{0}$ to $\ketF{N}$.

\subsection{Limiting Behaviors of the HPT - The Flat Bloch Sphere Approximation}\label{sub:HPTLimits}

The HPT is frequently treated in the limit of large $N$, compared to either the number of qubits in the $\ketQ{1}$ state (on the qubit side), or, equivalently, the index $k$ of the Fock state $\ketF{k}$ (on the bosonic side). In this limit, the operators approach the simpler form
\begin{subequations}
\begin{align}
    \mathbf{J}_+ &\HPTnp \sqrt{N} \, a^\dagger \\
    \mathbf{J}_- &\HPTnp \sqrt{N} \, a \\
    \mathbf{J}_x &\HPTnp \frac{\sqrt{N}}{2} \left( a^\dagger + a \right) = \sqrt{J} \, x \\
    \mathbf{J}_y &\HPTnp -i\frac{\sqrt{N}}{2} \left( a^\dagger - a \right) = -\sqrt{J} \, p
\end{align}
\end{subequations}
where $x = (a^\dagger + a)/\sqrt{2}$ and $p = i(a^\dagger - a)/\sqrt{2}$ are the usual degrees of freedom of a harmonic oscillator, with easily calculable corrections to these correspondences suppressed by powers of $k/N$. The qubit and boson systems are thus equivalent at zeroth order in $k/N$, with the $J_\pm$ operators corresponding precisely (up to a rescaling) to the bosonic raising and lowering operators. We label this limit as the ``south pole,'' since this will be this region's position on the Bloch sphere in our conventions.

This limit has a straightforward geometric interpretation: the bottom of the (Bloch) sphere, i.e. the vicinity of the south pole, is flat. This is most easily seen for coherent states, which are characterized by a single direction on the Bloch sphere (in the qubit case) or a position in the $x$-$p$ plane (in the boson case). Then, in our $k \ll N$ limit, we show in Section~\ref{sub:CoherentStates} that we have the correspondence\footnote{Although note that, as we discuss in that section, this is a correspondence based on preparation, rather than one based on number basis.}
\begin{align}
    \ketB{\theta, \phi}^{\otimes N} = \left( \sin\frac{\theta}{2} \ketQ{0} + e^{-i\phi} \cos\frac{\theta}{2} \ketQ{1} \right)^{\otimes N} &\HPTnp \ketC{\sqrt{N} e^{-i\phi} \cos\frac{\theta}{2}} \label{eq:HPTCoherentStates}
\end{align}
where we have introduced the notation $\ketB{\theta,\phi}$ for qubit states given by their Bloch vector's spherical coordinates (with $\theta$ measured from the north pole as usual), and $\ketC{\alpha}$ for bosonic coherent states with index $\alpha$. Note that, in the $\pi-\theta \ll 1$ limit, the coordinate of the right hand side is precisely that of of the $(\theta,\phi)$ position of a radius-$\sqrt{N}/2$ sphere with south pole at the origin of the complex plane, up to a sign resulting from our which-pole convention.

For our purposes, however, a different limit will often be more useful: that of (almost) exactly half of the qubits being in each of $\ket{0}$ and $\ket{1}$, i.e. of $|M| \ll J$, corresponding to states supported only on Fock states $\ket{k}$ with $k$ close to $N/2$. In this limit, we instead have
\begin{subequations}
\begin{align}
    \mathbf{J}_+ &\HPTeq \sqrt{J} \, a^\dagger \label{eq:HPTEquatorial1} \\
    \mathbf{J}_- &\HPTeq \sqrt{J} \, a \label{eq:HPTEquatorial2} \\
    \mathbf{J}_x &\HPTeq \frac{\sqrt{N}}{2\sqrt{2}} \left( a^\dagger + a \right) = \sqrt{\frac{J}{2}} \, x \label{eq:HPTEquatorial3} \\
    \mathbf{J}_y &\HPTeq -i\frac{\sqrt{N}}{2\sqrt{2}} \left( a^\dagger - a \right) = -\sqrt{\frac{J}{2}} \, p, \label{eq:HPTEquatorial4}
\end{align}
\end{subequations}
different from the few-excitation case only by a factor of $\sqrt{2}$. This again has a straightforward geometric interpretation: a narrow band of the Bloch sphere around the equator is equivalent to a narrow annulus in the bosonic plane. 

One can of course construct analogous limits near any other fixed latitude, with both the previous $M \approx -J$ and this $M \approx 0$ limits as special cases, e.g.
\begin{align}
    \mathbf{J}_x &\HPTapp \frac{\sqrt{N - \expect{n}}}{2} \left( a^\dagger + a \right), \label{eq:ApproxHPTJxx}
\end{align}
where $J+\expect{\mathbf{J}_z} \HPT \expect{n}$ sets the latitude where the approximation applies. This general form will be the basis for our definition of coherent states below.

\subsection{The HPT for Qudits and for Multiple Cavities}\label{sub:QuditHPT}

The HPT can also be applied to qudits (i.e. to collections of $d$-level systems), with the correspondence now between symmetric states of $N$ qudits and $m = d-1$ distinct bosonic modes~\cite{Kurucz:2009wkc}. This is most easily defined via the states:
\begin{align}
    \ketD{N-\sum_{k=1}^{d-1}N_k, N_1, N_2, ..., N_{d-1}} \HPT \prod_{k=1}^{d-1} \ketF{N_k}
\end{align}
where $\ketD{N_0, N_1, ..., N_{d-1}}$ is the symmetric state with $N_0$ qudits in their lowest (e.g. ground) level, $N_1$ qudits in the next (e.g. first excited) level, and so on. The occupation of the $k$'th bosonic mode thus corresponds to the number of qudits excited into the $k$'th level.

Following Ref.~\cite{Kurucz:2009wkc}, one can then define counting operators (analogous to $\mathbf{J}_z$ for $d=2$),
\begin{align}
    \Sigma_{k,k}\ketD{...,N_k,...} &= N_k\ketD{...,N_k,...}
\end{align}
as well as swapping operators (analogous to $\mathbf{J}_\pm$ for $d=2$),
\begin{align}
    \Sigma_{k,\ell}\ketD{...,N_k,...,N_\ell,...} &= \sqrt{(N_k+1)N_\ell} \ketD{...,N_k+1,...,N_\ell-1,...},
\end{align}
for $k \neq \ell$. It is then straightforward to show that, for $k>0$,
\begin{subequations}
\begin{align}
    \Sigma_{0,0} &\HPT N - \sum_{k=1}^{d-1} a_k^\dagger a_k = N - \sum_{k=1}^{d-1} \hat{n}_k \\
    \Sigma_{0,k} &\HPT \sqrt{ N - \sum_{\ell=1}^{d-1} \hat{n}_\ell } \, a_k \\
    \Sigma_{k,0} &\HPT a_k^\dagger \sqrt{ N - \sum_{\ell=1}^{d-1} \hat{n}_\ell } \\
    \Sigma_{k,\ell} &\HPT a_k^\dagger a_\ell.
\end{align}
\end{subequations}

In this work we will not use this generalized HPT explicitly, but it is conceptually valuable as the intuition underlying Section~\ref{sec:NmScaling}.

\section{Examples of the HPT Between States}\label{sec:examples}

We now illustrate the HPT on a series of physically important states, showing the bosonic Wigner function alongside the spin Wigner function on the Bloch sphere in each case. A Wigner function is a quasiprobability distribution over the corresponding phase space which acts as a convenient representation of states in the phase space of two non-commuting quadratures. For bosons, its marginals give the usual measurement probabilities, and negative values are a signature of nonclassical behavior; both of these properties are disrupted for qubits. A brief technical description of Wigner functions for bosons and (symmetric) qubit states, along with our conventions for them, is presented in Appendix~\ref{app:WignerFunctions}. For each of the following example states we will sketch the Wigner function, and in some cases, where the states wrap the Bloch sphere, we will additionally plot the spin Wigner function via an equirectangular projection for better visibility of certain features.

\subsection{The South-Pole State $\xleftrightarrow{\ \ \ }$ The Bosonic Vacuum} \label{sub:VacuumHPT}

\begin{figure}[htbp]
    \centering
    \includegraphics[valign=m, width=0.3\textwidth]{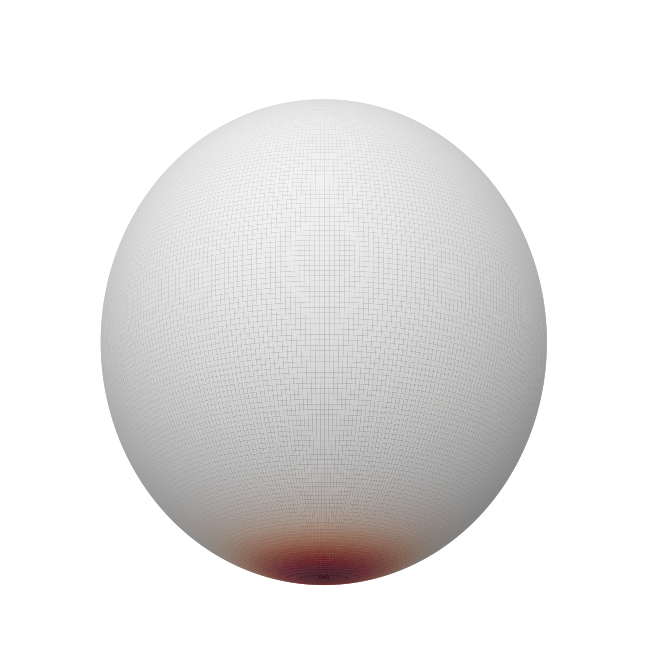}
    {\Large $\HPT$}\qquad
    \includegraphics[valign=m, width=0.3\textwidth]{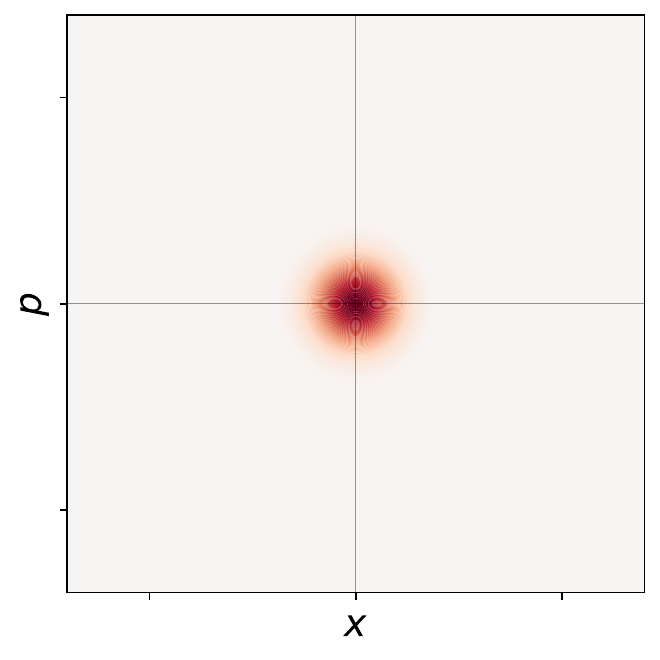}
    \caption{Wigner distributions for the south-pole state $\ketQ{0}^{\otimes N}$ of $N$ qubits (left) and for the bosonic vacuum $\ketF{0}$ (right); these map to one another under the HPT. The qubit Wigner distribution is approximately gaussian; this is exact on the boson side. We discuss this mapping in Section~\ref{sub:VacuumHPT}, while Appendix~\ref{app:WignerFunctions} presents our Wigner function conventions.}
    \label{fig:vacuum}
\end{figure}

The simplest state on the bosonic side is the vacuum $\ketF{0}$, the unique state annihilated by $a$. Under the HPT defined above the vacuum maps to the state which is annihilated by $\mathbf{J}_-$. This is a state  with zero excitations in which all $N$ qubits point down along $\hat z$. In our notation, this is the fully polarized state at the south pole of the Bloch sphere, $\ketD{J,-J}=\ketQ{0}^{\otimes N} \HPT \ketF{0}$. Both the bosonic and qubit Wigner functions are localized Gaussian-like peaks: at the origin of phase space on the bosonic side and at the south pole on the qubit side (Fig.~\ref{fig:vacuum}). The vacuum  will serve as a starting point for the rest of this section, where we will show examples achieved by acting on this state with various operators on both sides of the HPT. 

\subsection{Dicke States $\xleftrightarrow{\ \ \ }$ Fock States} \label{sub:DickeFockHPT}

A ladder of states can be built from the vacuum by acting on it with raising or creation operators. On the bosonic side Fock states are defined as
\begin{equation}
    \ketF{n}=\frac{\left(a^\dagger\right)^n}{\sqrt{n!}}\ketF{0}. 
\end{equation}
Fock states $\ketF{n}$ are eigenstates of the number operator $\hat n = a^\dagger a$ with eigenvalue $n$. They are the energy eigenstates of the harmonic oscillator and form the natural basis of $\Hbos$. The Wigner function of $\ketF{n}$ is azimuthally symmetric in phase space, with $n$ concentric rings of alternating sign.

\begin{figure}[t]
    \centering
    \includegraphics[valign=m, width=0.3\textwidth]{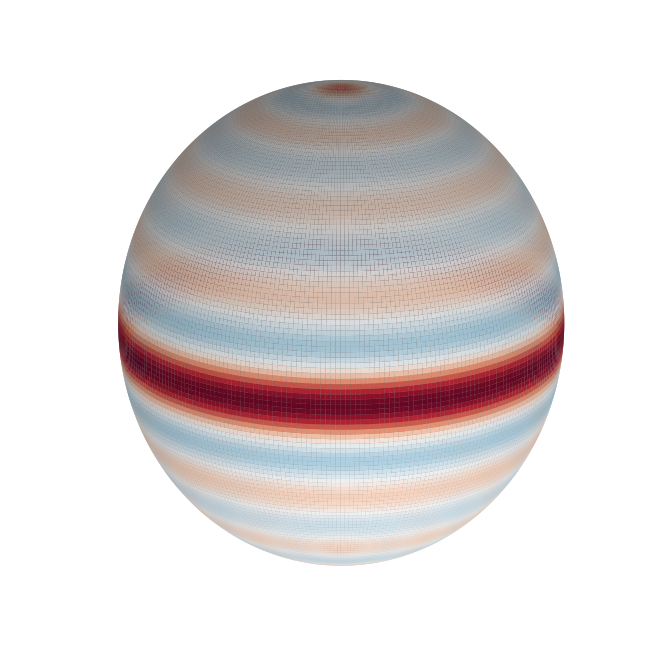}
    {\Large $\HPT$}\qquad
    \includegraphics[valign=m, width=0.3\textwidth]{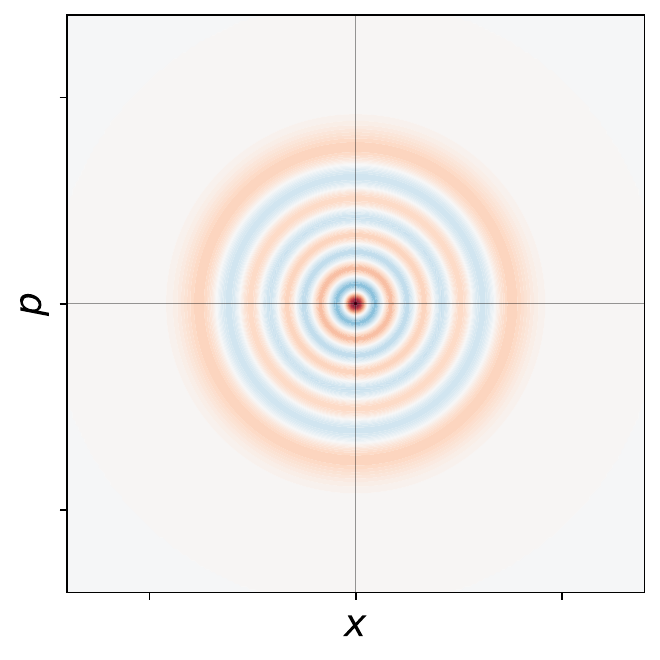}
    \caption{Wigner distributions for the permutation-symmetric, half-up state $\ketD{J=N/2,M=0}$ of $N=16$ qubits (left) and for the bosonic  state of $8$ photons $\ketF{N/2}$ (right); these map to one another under the HPT. Both states have definite excitation numbers, and both Wigner functions exhibit circularly symmetric rings of alternating sign. We discuss this mapping in Section~\ref{sub:DickeFockHPT}, while Appendix~\ref{app:WignerFunctions} presents our Wigner function conventions.}
    \label{fig:fock-dicke}
\end{figure}

Under the HPT,
\begin{equation}
     \ketD{J,n-J} \HPT \ketF{n} \label{eq:HPTDickeFock},
\end{equation} 
i.e. a Fock state with $n$ excitations maps to the Dicke state in which $n$ of the $N$ qubits have been flipped from $\ketQ{0}$ to $\ketQ{1}$, symmetrized over all $\binom{N}{n}$ arrangements. Just as the Fock state has a definite photon number and no preferred phase, the Dicke state has a definite $M$ eigenvalue and no preferred azimuthal orientation on the Bloch sphere. This shared azimuthal symmetry is visible in the Wigner functions: the bosonic ring structure maps to equatorial bands on the sphere
(Fig.~\ref{fig:fock-dicke}).

\subsection{Coherent States}\label{sub:CoherentStates}

Coherent states of both qubits and bosons can be obtained by displacing or rotating the vacuum. Indeed, both types are instances of the generalized coherent states defined in Ref.~\cite{Perelomov1972}: states obtained by the action of a Lie group on a fiducial state. A bosonic coherent state is thus obtained by displacing the vacuum,
\begin{equation}
    \ketF{\alpha} = D(\alpha)\ketF{0}
    = e^{\alpha a^\dagger - \alpha^* a}\ketF{0}
    = e^{-|\alpha|^2/2}\sum_{n=0}^{\infty}
      \frac{\alpha^n}{\sqrt{n!}}\ketF{n},
\end{equation}
producing a Poissonian superposition of Fock states with mean occupation $\expect{\hat{n}} = |\alpha|^2$. The amplitude $|\alpha|$ sets the distance from the origin in phase space while $\arg(\alpha)$ sets the orientation.

\begin{figure}[t]
    \centering
    \includegraphics[valign=m, width=0.3\textwidth]{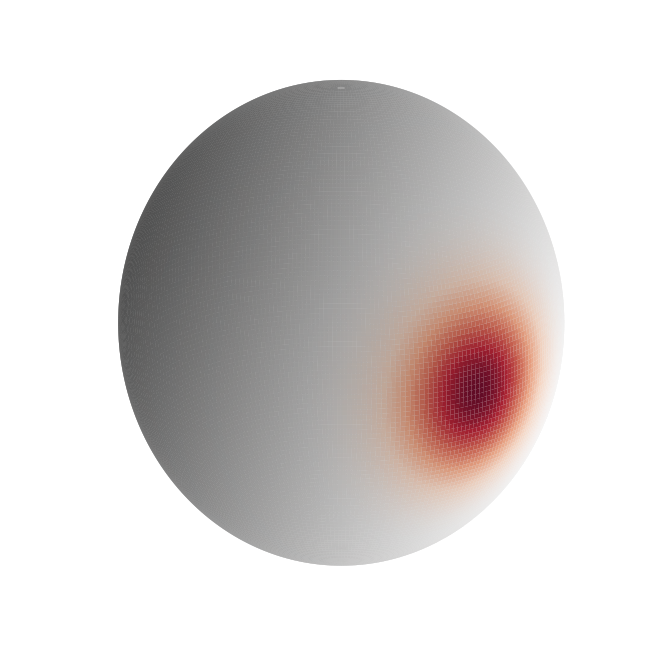}
    {\Large $\HPT$}\qquad
    \includegraphics[valign=m, width=0.3\textwidth]{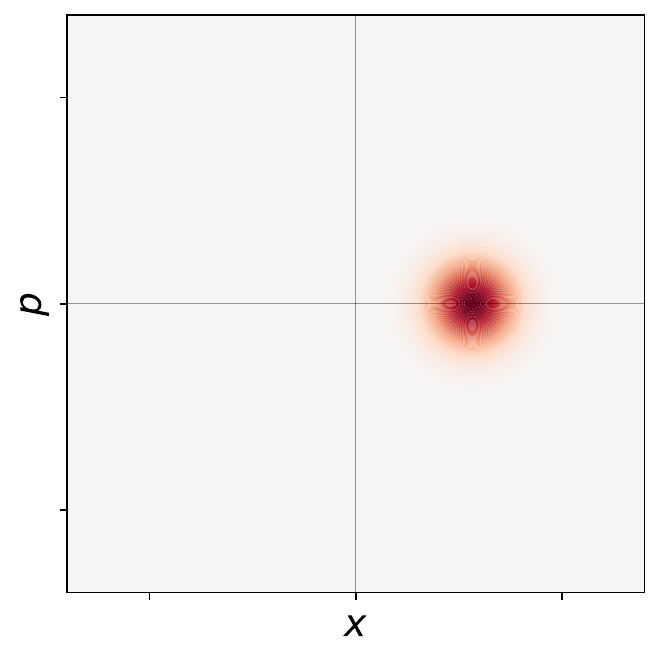}
    \caption{Wigner distributions for an example displaced coherent state of qubits (left) and of a bosonic mode (right); these map to one another under the HPT. As in the vacuum (coherent) case, the qubit distribution is approximately gaussian while gaussianity is exact for the bosonic Wigner function. We discuss this mapping in Section~\ref{sub:CoherentStates}, while Appendix~\ref{app:WignerFunctions} presents our Wigner function conventions.}
    \label{fig:coherent}
\end{figure}

The analog on the qubit side is the spin coherent state, obtained by rotating the south-pole state:
\begin{equation}
    \ketD{\theta,\varphi} = R(\theta,\varphi)\ketD{J,-J},
\end{equation}
where $R(\theta,\varphi) = e^{-i(\pi-\theta)(\mathbf{J}_x\sin\varphi- \mathbf{J}_y\cos\varphi)}$ is an $SU(2)$ rotation. Expanding in the Dicke basis  gives a binomial superposition,
\begin{equation}
    \ketD{\theta,\varphi}
    = \sum_{M=-J}^{J}\binom{N}{J-M}^{1/2}
      \cos^{J+M}\!\tfrac{\theta}{2}\;
      \sin^{J-M}\!\tfrac{\theta}{2}\;
      e^{-i(J+M)\varphi}\;\ketD{J,M}.
\end{equation}
The polar angle $\theta$ plays the role of $|\alpha|$: it controls the spread over the Dicke basis, with $\theta=\pi/2$ (the equator) maximizing this spread, analogous to large $|\alpha|$. The azimuthal angle $-\varphi$ plays the role of $\arg(\alpha)$. On both sides, the Wigner function is a localized peak (Fig.~\ref{fig:coherent}).

Note that we are defining the HPT correspondence between states based on the approximate HPT correspondence between the displacements/rotations above, rather than based on their decomposition in the Dicke/Fock basis. Integrating both sides of Eq.~\eqref{eq:ApproxHPTJxx} to achieve the equatorial coherent state and its corresponding HPT state, for example, gives
\begin{align}
    e^{i\int_0^{\pi/2} d\tau \, \mathbf{J}_x} \ketQ{0}^{\otimes N} \HPT e^{\frac{i}{2}\int_0^{\pi/2} d\tau \, \sqrt{N-\expect{n}} \, (a^\dagger + a)} \ketB{0}
\end{align}
which give
\begin{align}
    \ketB{\theta = \frac{\pi}{2}, \phi=-\frac{\pi}{2}}^{\otimes N} &\HPT \ketC{\alpha = i\sqrt{\frac{N}{2}}},
\end{align}
matching Eq.~\eqref{eq:HPTCoherentStates}. This is not the same as decomposing each state in the Dicke/Fock basis and then matching these via Eq.~\eqref{eq:HPTDickeFock}. Although the latter reproduces the correct expectation values of $\hat{n}$ and $\mathbf{J}_z$, it would also give equal variances of $\hat{n}$ and $\mathbf{J}_z$, whereas the variances of the actual coherent states are different: qubit coherent states have a number variance of only $N/4$, whereas for bosonic coherent states it is $N/2$. This is why we choose the preparation-based definition rather than Dicke/Fock-based definition; the latter would lead to coherent states on one side corresponding to slightly squeezed states on the other side, introducing factors of 2 throughout our discussion. This does not affect anything parametrically however.

Two specific coherent states will be particularly useful to us later:
\begin{subequations}
\begin{align}
    \ketQ{+}^{\otimes N} = \pfrac{\ketQ{0}+\ketQ{1}}{\sqrt{2}}^{\otimes N} = \ketB{\theta = \frac{\pi}{2},\phi=0}^{\otimes N} &\HPT \ketC{\alpha = \sqrt{\frac{N}{2}}} \\
    \ketQ{-}^{\otimes N} = \pfrac{\ketQ{0}-\ketQ{1}}{\sqrt{2}}^{\otimes N} = \ketB{\theta = \frac{\pi}{2},\phi=\pi}^{\otimes N} &\HPT \ketC{\alpha = -\sqrt{\frac{N}{2}}}.
\end{align}
\end{subequations}

\subsection{Squeezed States}\label{sub:SqueezedStates}

\begin{figure}[t]
    \centering
    \includegraphics[valign=m, width=0.3\textwidth]{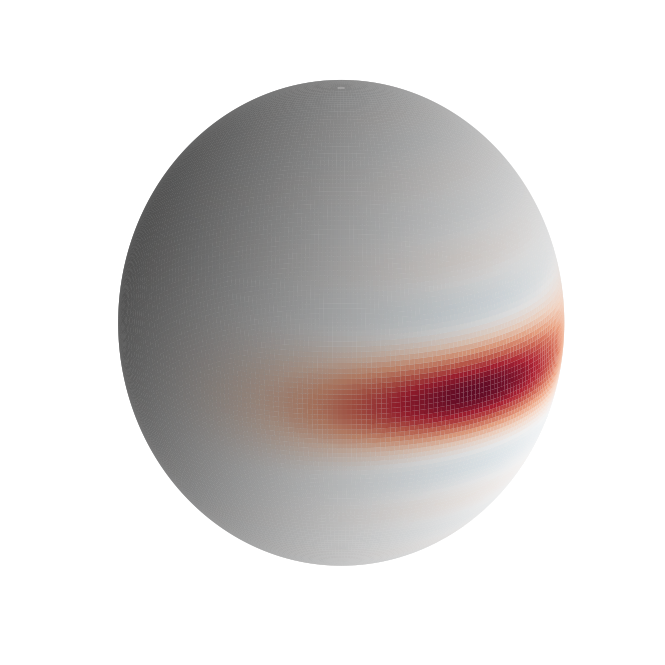}
    {\Large $\HPT$}\qquad
    \begin{minipage}[c]{0.30\textwidth}
    \centering
    \includegraphics[width=0.8\linewidth]{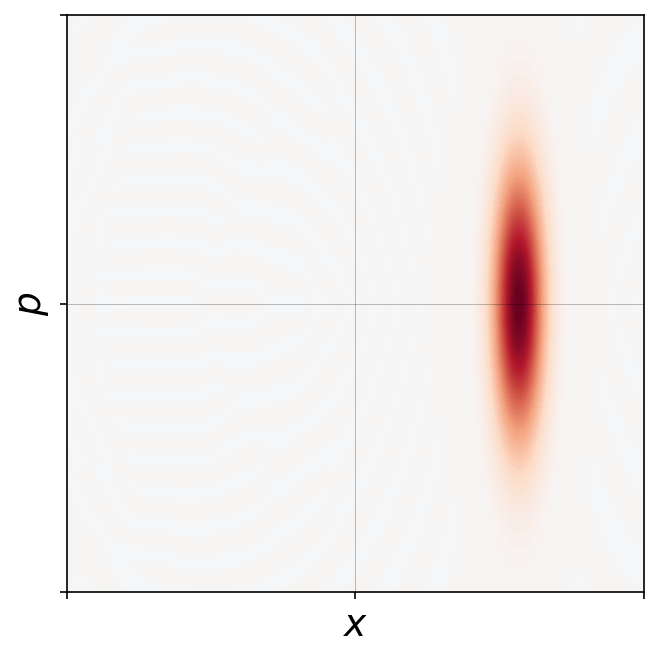} \\ \vspace{1mm} ~~\,OR \\
    \vspace{3.2mm}
    \includegraphics[width=0.8\linewidth]{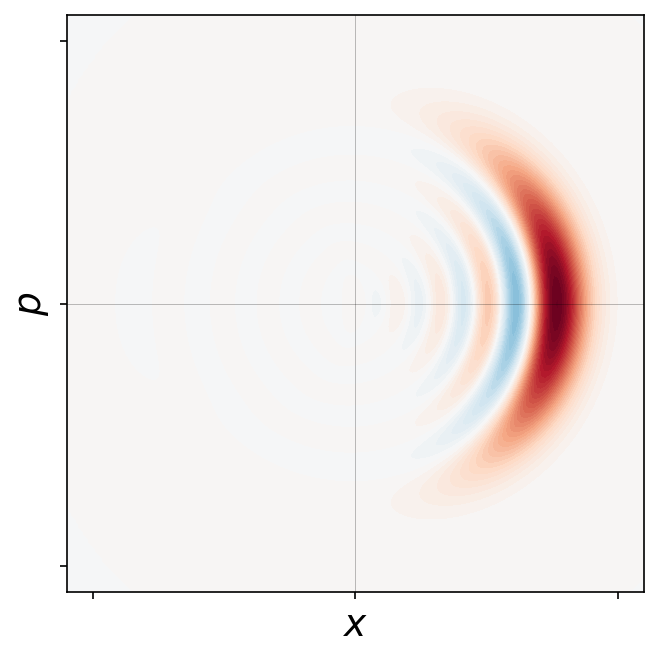}
    \end{minipage}
    \caption{Wigner distributions for an example spin-squeezed (along $\mathbf{J}_z$) state of qubits (left) and for two types of squeezing of a bosonic mode (right): $x$ quadrature squeezing (upper-right) and number-squeezing (lower-right). The HPT map in this case is somewhat more ambiguous that in the rest of our examples; see the discussion in Sections~\ref{sub:SqueezedStates} and~\ref{subsub:SqueezedProtocols}. Appendix~\ref{app:WignerFunctions} presents our Wigner function conventions.}
    \label{fig:squeezed}
\end{figure}

Squeezed states redistribute the fluctuations of a coherent state, reducing the uncertainty in one direction at the expense of increased uncertainty in another. On the bosonic side, a displaced squeezed vacuum can be written as
\begin{equation}
    \ket{\alpha,r}
    = D(\alpha)S(r)\ketF{0},
    \qquad
    S(r)=\exp\!\left[\frac{r}{2}
    \left(a^2-(a^\dagger)^2\right)\right].
\end{equation}
For real $r>0$, this state has
\begin{equation}
    \Var(x)=\frac{e^{-2r}}{2},
    \qquad
    \Var(p)=\frac{e^{2r}}{2}.
\end{equation}
Its Wigner function is an elliptical Gaussian, compressed along $x$ and elongated along $p$. Taking $\alpha$ real and positive places this ellipse on the positive $x$ axis, as illustrated in the upper-right panel of Fig.~\ref{fig:squeezed}.

The analogous construction on the qubit side is a spin-squeezed state~\cite{Kitagawa1993}, up to a subtlety we return to below. An equatorial spin coherent state polarized along $x$ has equal transverse fluctuations, $\Var(\mathbf{J}_y)=\Var(\mathbf{J}_z)=J/2$. A state that remains approximately polarized along $x$ can instead have reduced fluctuations along $z$ and increased fluctuations along $y$. Its Wigner function is then narrow in latitude and extended in azimuth, as in the left-hand panel of Fig.~\ref{fig:squeezed}. However, under the exact HPT,
\begin{equation}
    \mathbf{J}_z \HPT \hat n-J, \qquad \Var(\mathbf{J}_z)=\Var(\hat n),
\end{equation}
so squeezing the distribution over Dicke levels corresponds directly to squeezing the distribution is occupation number, rather than quadrature.

This number squeezing is closely related to the quadrature squeezing described above, but the two are not identical. For a state localized near the positive $x$ axis, we can expand
\begin{equation}
    \hat n=\frac{x^2+p^2-1}{2}, \qquad \delta\hat n\simeq\expect{x}\,\delta x,
\end{equation}
provided the quadratic fluctuations can be neglected. In this limit, squeezing $x$ also squeezes the occupation number. For a state with greater extent in the tangential direction, however, the curvature of the contours of constant occupation becomes relevant, and thus the quadrature-squeezed bosonic state becomes different from the number-squeezed bosonic state. A number-squeezed state will then have a curved, rather than elliptical, Wigner function, as illustrated in the lower-right panel of Fig.~\ref{fig:squeezed}. The two bosonic pictures agree locally when this curvature is negligible.

\subsection{The GHZ State $\xleftrightarrow{\ \ \ }$ The Cat State} \label{sub:GHZCatHPT}

\begin{figure}[t]
    \centering
    \begin{minipage}[c]{0.30\textwidth}
    \centering
    \includegraphics[width=0.85\linewidth]{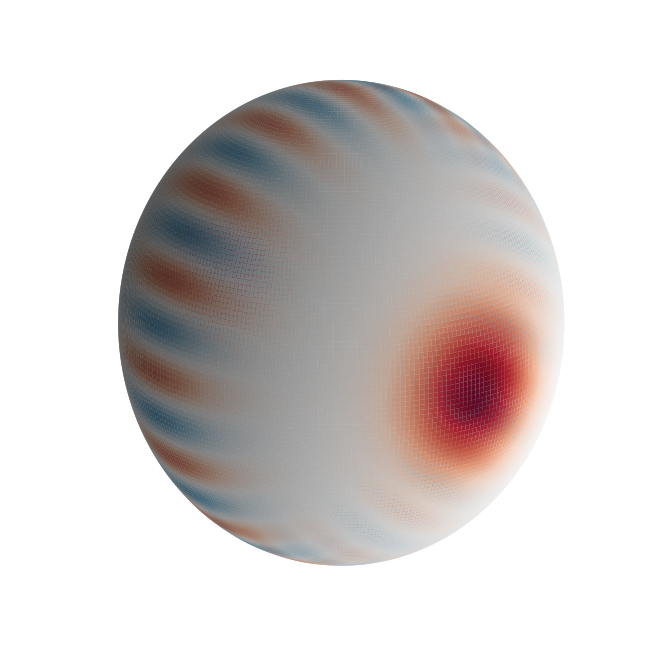}
    \includegraphics[width=\linewidth]{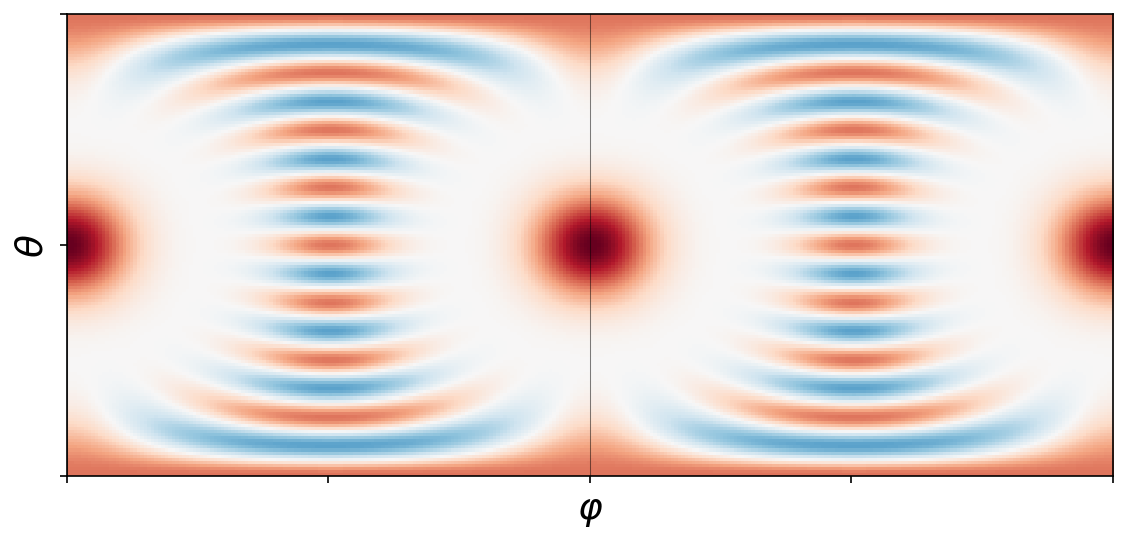}
    \end{minipage}
    \hspace{0.012\textwidth}
    \begin{minipage}[c]{0.14\textwidth}
    \centering
    {\Large $\HPT$}
    \end{minipage}
    \hspace{0.015\textwidth}
    \begin{minipage}[c]{0.30\textwidth}
    \centering
    \includegraphics[width=\linewidth]{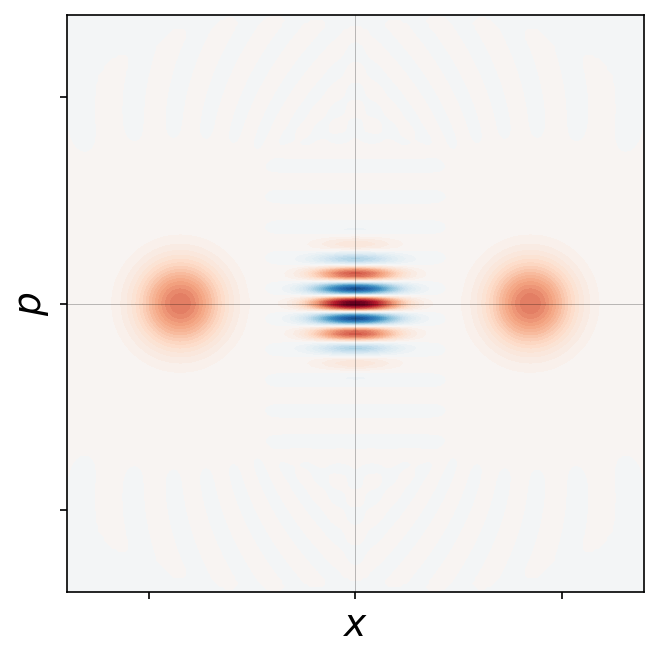}\vspace{-7mm}
    \end{minipage}
    \caption{Wigner distributions for a qubit GHZ(-2) state aligned with the $\mathbf{J}_x$ axis (left) and for a bosonic (2-legged) cat state along the $x$ axis (right); these map to one another under the HPT. For the qubit Wigner function, we additionally include a plot in spherical coordinates $(\theta,\phi)$ to show the entire spherical surface. Both states have two primary lobes at the positions of their component coherent states, as well as interference fringes of alternating sign in between. We discuss this mapping in Section~\ref{sub:GHZCatHPT}, while Appendix~\ref{app:WignerFunctions} presents our Wigner function conventions.}
    \label{fig:cat2-ghz}
\end{figure}

The 2-legged cat state can be viewed in two equivalent ways. First, it is a superposition of two coherent states with opposite phases:
\begin{equation}\label{eq:cat2-superposition}
    \catN{2} \propto \ketC{\alpha} + \ketC{-\alpha}.
\end{equation}
Second, it arises from projecting a coherent state onto the subspace of even Fock numbers. The projector onto even occupation is
\begin{equation}
    \Pi_{\rm even} = \frac{1}{2}\bigl(1 + e^{i\pi\hat n}\bigr)
    = \frac{1}{2}\bigl(1 + (-1)^{\hat n}\bigr),
\end{equation}
so that
\begin{equation}\label{eq:cat2-projection}
    \Pi_{\rm even}\ketC{\alpha}
    = \frac{1}{2}\bigl(\ketC{\alpha} + \ketC{-\alpha}\bigr)
    \propto \sum_{n \equiv 0\;(\mathrm{mod}\;2)} \frac{\alpha^n}{\sqrt{n!}}\ketF{n}.
\end{equation}
The rotation $\alpha \to -\alpha$ is generated by $e^{i\pi\hat n}$, so the two pictures are equivalent: superposing two antipodal coherent states is the same as selecting every other Fock component. On the qubit side, the same construction applies with $\hat n\to  J+\mathbf{J}_z$ and the parity operator $(-1)^{\hat n}\to e^{i\pi(J+\mathbf{J}_z)} = Z^{\otimes N}$, where $Z$ is $1$ for $\ketQ{0}$ and $-1$ for $\ketQ{1}$. The role of the coherent state $\ketF{\alpha}$ is played by the equatorial spin coherent state $\ketD{\pi/2,0}=\ketQ{+}^{\otimes N}$. Projecting onto the $+1$ eigenspace of $Z^{\otimes N}$ thus gives the GHZ state along the $x$ axis
\begin{equation}
    \GHZxN{} \propto \Pi_{Z=+1}\ketQ{+}^{\otimes N}
    \propto \sum_{\substack{M=-J \\ J+M \equiv 0\;(\mathrm{mod}\;2)}}^{J}
    \binom{N}{J-M}^{1/2}\ketD{J,M}. \label{eq:GHZinFockBasis}
\end{equation}
Note that these coefficients peak at $|M| \ll J$, rather than at the extreme values $|M| = J$. This is the result of our choice of axes for the GHZ state ($x$) and for our basis ($z$): although a GHZ state oriented along the $z$ axis would have support only at $|M| = J$, we are instead showing the GHZ state oriented along the $x$ axis, which has support at all $M$ such that $M+J$ is even, but primarily those with $|M| \lesssim \sqrt{J}$; see Fig.~\ref{fig:cat2-ghz}.

\subsection{The GHZ-4 State $\xleftrightarrow{\ \ \ }$ The 4-Legged Cat State}\label{sub:Z4States}

The construction generalizes to any $\mathbb{Z}_k$ symmetry. We focus on $k=4$ here, which will serve as our primary example in Section~\ref{sec:DetectionUsingGHZ4}, but generalizing the following to other $k$ values is trivial.

The 4-legged cat state can again be described in two equivalent ways: As a superposition of four coherent states, it is
\begin{equation}\label{eq:cat4-superposition}
    \catN{4} \propto \ketC{\alpha} + \ketC{-\alpha} + \ketC{i\alpha} + \ketC{-i\alpha}.
\end{equation}
As a $\mathbb{Z}_4$ projection, it instead selects every fourth Fock component, i.e.
\begin{equation}\label{eq:cat4-projection}
    \Pi_{0\,\mathrm{mod}\,4}\ketC{\alpha} = \frac{1}{4}\sum_{l=0}^{3} e^{i\pi l\hat n/2}\ketC{\alpha} \propto \sum_{n\equiv 0\,({\rm mod}\;4)} \frac{\alpha^n}{\sqrt{n!}}\ketF{n}.
\end{equation}
The four rotations $\alpha\to i^l\alpha$ in phase space are generated by $e^{i\pi\hat n/2}$, so the superposition and projection pictures again coincide.

\begin{figure}[t]
    \centering
    \begin{minipage}[c]{0.30\textwidth}
    \centering
    \includegraphics[width=0.85\linewidth]{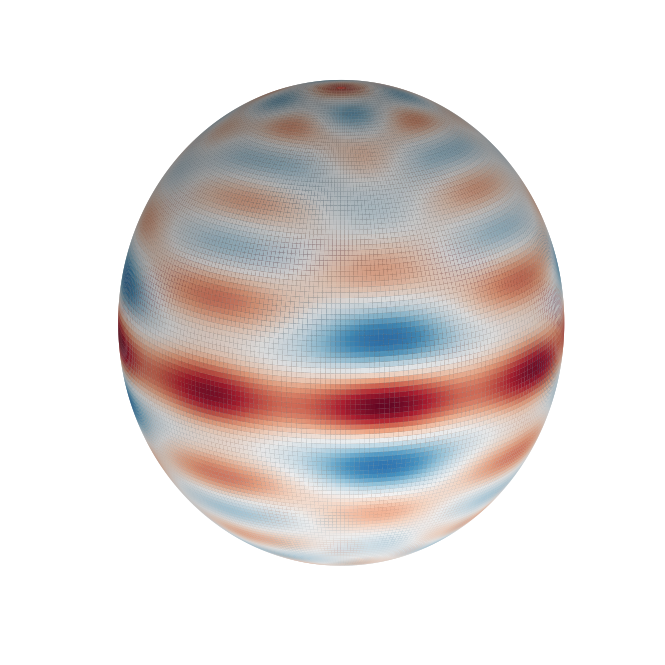}
    \includegraphics[width=\linewidth]{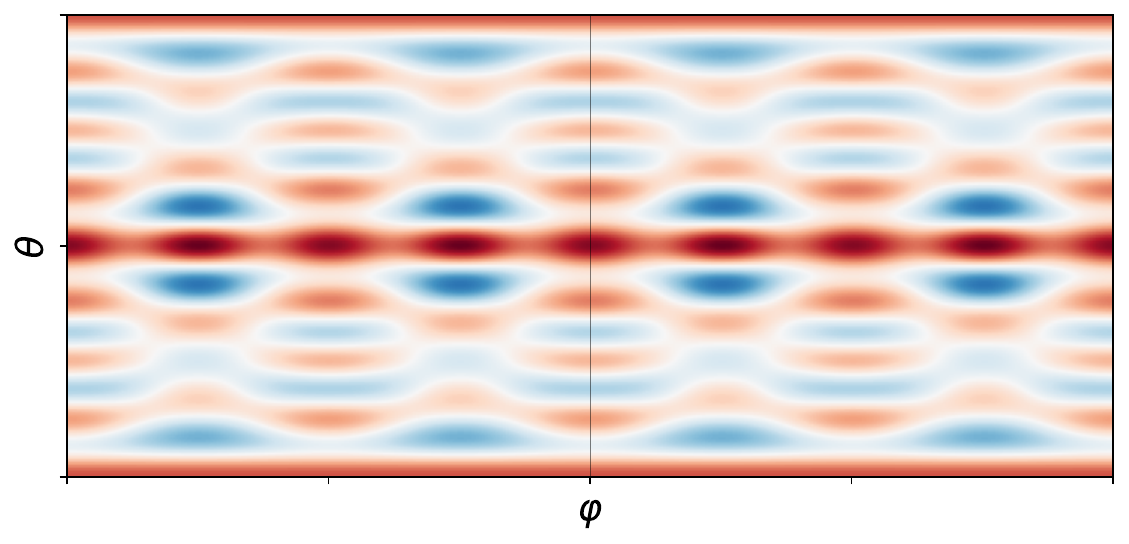}
    \end{minipage}
    \hspace{0.012\textwidth}
    \begin{minipage}[c]{0.14\textwidth}
    \centering
    {\Large $\HPT$}
    \end{minipage}
    \hspace{0.015\textwidth}
    \begin{minipage}[c]{0.30\textwidth}
    \centering
    \includegraphics[width=\linewidth]{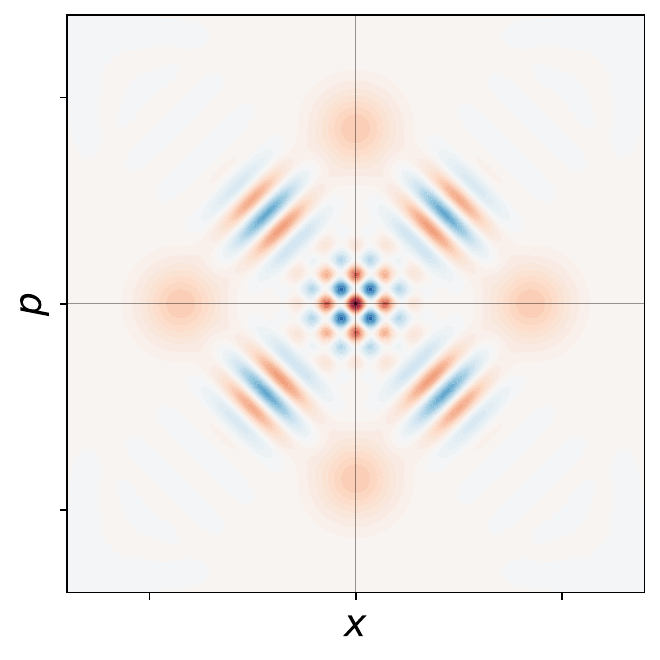}\vspace{-7mm}
    \end{minipage}
    \caption{Wigner distributions for a qubit GHZ-4 state aligned with the $\mathbf{J}_x$ and $\mathbf{J}_y$ axes (left) and for a bosonic 4-legged cat state along the $x$ and $p$ axes (right); these map to one another under the HPT. For the qubit Wigner function, we additionally include a plot in spherical coordinates $(\theta,\phi)$ to show the entire spherical surface. Although the four-fold symmetry and the four constituent coherent states are still visible on both sides, the interference fringes on the qubit side are much more pronounced than in the GHZ-2 case; see Fig.~\ref{fig:cat2-ghz}. We discuss this mapping in Section~\ref{sub:Z4States}, while Appendix~\ref{app:WignerFunctions} presents our Wigner function conventions.}
    \label{fig:cat4-ghz}
\end{figure}

On the qubit side, the $\mathbb{Z}_4$ generator becomes $e^{-i\pi(J+\mathbf{J}_z)/2}$, and projecting the equatorial spin coherent state onto its $+1$ eigenspace gives what we will call the ``GHZ-4'' state,
\begin{equation}
    \GHZxN{4} \propto \Pi_{\mathbb{Z}_4}\ketQ{+}^{\otimes N} \propto \sum_{\substack{M=-J\\J+M\equiv 0\;(\mathrm{mod}\;4)}}^{J} \binom{N}{J-M}^{1/2}\ketD{J,M}. \label{eq:GHZ4inFockBasis}
\end{equation}
For $N=2J$ qubits, this selects Dicke states at $M$ values spaced by 4. The four-fold symmetry is visible in both the phase-space and Bloch-sphere Wigner functions (Fig.~\ref{fig:cat4-ghz}).

\section{Examples of the HPT Between Sensing Protocols}\label{sec:implications}

Having described the HPT correspondence between an array of states, we now discuss its applicability to a variety of experimental techniques, emphasizing the correspondence between certain qubit- and boson-based approaches. We then present a novel class of easily-producible qubit states with improved sensitivity to light dark matter.

\subsection{Signal Hamiltonians}
Throughout this section, we consider two analogous interactions: On the qubit side, we consider interactions that induce rotations; on the boson side we instead consider displacements. These are the unique interactions to appear at lowest order in the rank of the interaction tensor (for qubits) or lowest order in creation or annihilation operators (for bosons), other than terms that introduce only an unobservable phase. They therefore appear across a wide range of dark matter searches, e.g. for dark photons \cite{Chen:2022quj, Agrawal:2023umy} or for axions \cite{Chen:2024aya, Brady:2022bus}.

For simplicity we will generally take the rotations to be $\exp(i\theta \mathbf{J}_y)$ about the $y$-axis, and likewise we focus on displacements in the $x$ direction, i.e. $\exp(i\beta p)$, with $\beta=-\theta \sqrt{J/2}$. In both cases, different contexts may or may not call for averaging over directions, but this does not qualitatively change the discussion below. These interactions are precisely analogous for all states near the equator\footnote{As usual, one can adapt this to other regions of the sphere by changing the $J$-dependent prefactors; see Section~\ref{sub:HPTLimits}} of the Bloch sphere, per Eqs.~\eqref{eq:HPTEquatorial1}--\eqref{eq:HPTEquatorial4}:
\begin{align}
    H = c\mathbf{J}_y &\HPTeq H = - c\sqrt{\frac{J}{2}}\, p
\end{align}
for evolution in a background field, with $c$ some constant dependent on the dark matter state and interaction strength. In scattering contexts, it is often most natural to write the analogous interactions as
\begin{align}
    \mathcal{L} \supset \lambda \mathcal{O}_{\rm DM}\mathbf{J}_- + {\rm h.c.} &\HPTeq \mathcal{L} \supset \lambda \sqrt{J} \mathcal{O}_{\rm DM} a + {\rm h.c.},
\end{align}
with $\mathcal{O}_{\rm DM}$ the dark matter part of the interaction operator, where the written-out terms correspond to dark matter removing one unit of angular momentum (one photon) while the Hermitian conjugates correspond to adding one unit of angular momentum (one photon); these are of course simply linear combinations of the $\mathbf{J}_{x}$ and $\mathbf{J}_y$ ($x$ and $p$) operators.

Rotations around the $z$-axis look slightly different, as Eq.~\eqref{eq:HPTGeneral1} maps $\mathbf{J}_z$ to $a^\dagger a = \hat{n}$ (up to a constant) rather than to any linear combination of $a$ and $a^\dagger$, but this does not introduce any fundamentally novel dynamics relative to the discussion below (although it does lead to different scaling with photon number; we return to this point in Section~\ref{sec:NmScaling}). Both $H \propto \mathbf{J}_z$ and $H \propto \hat{n}$ are effectively just rotations around the vacuum (i.e. the polar axis or the origin).

As an instructive example, we consider the action of a rotation/displacement pair, $H = c\mathbf{J}_x \longleftrightarrow H = c\sqrt{J/2}\, x$, on the GHZ-2/cat-2 state. Applying these Hamiltonians for some time $\tau$ gives unitary evolution
\begin{align}
    U = e^{-i c \tau \mathbf{J}_x } \HPTeq U = e^{-i c\sqrt{J/2}\, \tau x},
\end{align}
corresponding to a rotation by angle $c\tau$ about the $x$ axis for qubits, or a displacement by $-c\sqrt{J}\, \tau/2$ along the ${\rm Im}(\alpha) \sim p$ axis for bosons. It may na{\"i}vely appear that these do not correspond to one another: after all, the displacement shifts both coherent state components of the cat state in the same direction away from the x-axis, whereas the rotation does not move the coherent components of the GHZ state at all; it merely introduces a phase between them. At lowest order, however, this is also true on the bosonic side: since the overlap between the coherent states is
\begin{align}
    \braopket{\alpha}{U}{\alpha} =  e^{-i\alpha c\sqrt{J}\, \tau/2} \braket{\alpha}{\alpha - i\frac{c\sqrt{J}\,\tau}{2}} = e^{-i\alpha c\sqrt{J}\, \tau} e^{-\frac{c^2 J \tau^2}{8}}
\end{align}
where we take $\alpha$ real since our cat state is along the $x$ axis, we see immediately that the only effect at $O(\alpha c\sqrt{J}\, \tau)$ is a phase, which, being linear in $\alpha$, will carry opposite sign between the $\ketC{\alpha}$ and $\ketC{-\alpha}$ components of the cat state. For the cat state corresponding to our GHZ state, with $\alpha = \pm \sqrt{J}$, this is a phase of $\mp cJ\tau$, precisely matching the phases acquired by the GHZ state components on the qubit side of the HPT.

This is especially evident if one instead considers the actions of these same Hamiltonians on the GHZ-4/cat-4 states. Now, the two additional coherent components of the GHZ-4 state (those along the $y$ axis) are rotated---one above the equator, one below---while the two new components of the cat-4 state ($\ketC{i\alpha}$ and $\ketC{-i\alpha}$) are shifted along the $p$ axis, both in one direction; since this shifts one component towards the origin and the other away, it matches the qubit picture cleanly. Both of these shifts, however, give state changes only at second order (i.e. $O(c^2\tau^2)$ for qubits or $O(\alpha^2 c^2 J \tau^2)$ for bosons).

\subsection{The HP Correspondence Between Known Measurement Techniques}

We now consider the HPT correspondence between a variety of existing experimental strategies, categorizing them by the HPT-analogous states that they employ. We begin with the Dicke state--Fock state correspondence, which is in many ways the simplest, before turning to coherent states and then superpositions thereof.

\subsubsection{Dicke States and Fock States}\label{subsub:DickeFockMeasurement}

There are, broadly speaking, two classes of Dicke or Fock state-based measurement techniques: those that begin in the vacuum and merely measure in a Dicke/Fock basis, and those that begin by preparing an excited Dicke/Fock state. We begin by briefly considering the former, which is far easier to implement and therefore more widespread today, before considering the difficult but potentially more powerful latter strategy.

In a Dicke state measurement starting from the vacuum, as in Ref.~\cite{Chen:2025tgj}, one begins in an all-down state $\ketD{J,-J} = \ketQ{0}^{\otimes N}$ and searches for a collective rotation around an axis that is uniformly randomized between measurements but fixed for the duration of a single measurement---$H \propto \cos\phi \, \mathbf{J}_x + \sin\phi \, \mathbf{J}_y$, restricting to the $x$-$y$ plane as a simple example---by measuring $\ket{W}\bra{W}$, where $\ket{W} = \ketD{J,-J+1}$ (or by measuring both $\mathbf{J}_z$ and $\mathbf{J}^2$, using the latter to reject most incoherent bit flips). The probability of seeing a signal after an integrated rotation of each qubit by angle $\theta$ is then $p \propto N\theta^2$, where the $N$ is a collective enhancement: the Hamiltonian for the collective rotational mode is enhanced by a factor of $\sqrt{N}$. This is the same signal probability as one would find by measuring each qubit independently, but the backgrounds are different, assuming the dominant background is single-qubit errors: the collective measurement gives an $N$-independent background rate, since one is effectively sensitive only to noise in a single mode, whereas the $N$ individual-qubit measurements can each experience individual errors and thus the total error rate is also proportional to $N$. Conversely, measurements of transverse angular momentum, e.g. $\mathbf{J}_x$, can detect only distribution broadening after averaging over $\phi$, parametrically worsening such a measurement's sensitivity scaling with the total number of measurements performed in the negligible-background regime.

The analogous Fock state measurement, as in Ref.~\cite{Dixit:2020ymh}, initiates a cavity in the vacuum and looks for random-direction displacement signals (e.g. due to a dark photon) by performing QND measurement of $\hat{n}$. As above, this measurement scales better with repeated measurement (at negligible background excitation rates) than, for example, measurements of $x$, as the initial state has zero variance in $\hat{n}$ but not in $x$. There is no direct analogue of scaling with $N$, however, since there is no number (of cavities, or of photons) with which sensitivity can scale; this will change below, and we discuss it further in Section~\ref{sec:NmScaling}.

We now turn to the advantages of preparing initial Dicke and Fock states that are not the vacuum for quantum sensing. The potential of Dicke states for precision measurements beyond the standard quantum limit is well-established~\cite{InterferometerCoherenceReview, He:2025ovo}, but they have more recently been identified as powerful target states for scattering experiments searching for feebly-coupled particles; this is the maximally-squeezed limit of Refs.~\cite{SuperradiantDetection1, SuperradiantDetection2, SuperradiantDetection3}. Both applications rely on two properties of (near-)equatorial Dicke states, i.e. $\ketD{J,M}$ with $|M| \ll J$. First, they are maximally spread in the transverse plane: combining $\mathbf{J}^2 = \mathbf{J}_x^2 + \mathbf{J}_y^2 + \mathbf{J}_z^2$ with $\expect{\mathbf{J}^2} = J(J+1)$, $\expect{\mathbf{J}_z^2} = M^2$, and the azimuthal symmetry $\expect{\mathbf{J}_x^2} = \expect{\mathbf{J}_y^2}$ gives
\begin{equation}\label{eq:DickeJx2}
    \expect{\mathbf{J}_x^2} = \expect{\mathbf{J}_y^2} = \frac{J(J+1) - M^2}{2}
    \;\xrightarrow[M=0]{}\; \frac{N(N+2)}{8},
\end{equation}
both quadratic in $N$. Because $\expect{\mathbf{J}_x} = 0$, this expectation value is also the quantum Fisher information for a rotation generated by $\mathbf{J}_x$, $F_Q = 4\expect{\mathbf{J}_x^2} \approx N^2/2$ at the equator---a factor of $\mathcal{O}(N)$ above the standard quantum limit. Second, $\ketD{J,M}$ is an eigenstate of $\mathbf{J}_z$, so a single exchanged quantum sends it to the orthogonal state $\ketD{J,M\pm 1}$. Even a single spin-flip is therefore detectable, with a rate enhanced by $\expect{\mathbf{J}_- \mathbf{J}_+} \approx N^2/4$~\cite{Bogorad:2026ggm}. The price of this effectiveness is difficult preparation: while quantum circuits to produce Dicke states with $O(N^2)$ \cite{DickeCircuit2019} or $O(N \log N)$ \cite{DickeCircuit2022} gates are known, reaching large $N$ is sufficiently difficult that Dicke state-based dark matter detection may not be practical for some time.

The analogous bosonic states are Fock states, and we consider two different classes of approaches to looking for light new physics using Fock states. First, Fock states of electromagnetic cavity modes \cite{Dixit:2020ymh, Agrawal:2023umy, Freiman:2025tse} can be used to search for dark photons or other electromagnetically-coupled dark matter via photon counting. In these searches, a cavity is initially prepared in a Fock state, and the occupation number is periodically read out using a quantum non-demolition (QND) measurement, achieved via a photon number-dependent shift in the transition frequency of a qubit in the cavity. If a dark matter particle's interaction with the detector results in a photon being added to the mode, this can therefore be observed as an upward step in the time series of the photon number. Notably, this upward step can be distinguished from the background of downward jumps caused by the mode's finite decay time; background jumps upward should not occur so long as the cavity temperature is well below the photon frequency. As in the qubit case, sensitivity is the result of a coherent enhancement---now caused simply by the $\sqrt{n+1}$ factor arising in transition amplitudes from applying a raising operator to $\ketF{n}$---combined with being in an eigenstate of $a^\dagger a$, the bosonic analogue of $\mathbf{J}_z$.

An entirely different approach to dark matter detection using Fock states is the quantum electron cyclotron, in which one or more electrons are held in the discrete cyclotron levels of a Penning trap \cite{Peil:1999nsg, Fan:2022uwu, Fan:2024mhm}. Since this is effectively a harmonic oscillator, it can be understood as an analogue of the electromagnetic cavity approach above, with the bosons now corresponding to quasiparticle excitations of the trap. An electron is initially prepared in a definite cyclotron mode, i.e. a Fock level, and its mode is regularly observed by measuring the frequency of a different mode of the electron's motion, e.g. the axial mode, which depends on the occupied cyclotron level; this forms a QND measurement entirely analogous to the cavity-qubit measurement above. Upward jumps of the Fock number can again be interpreted as signals of energy deposition, e.g from dark photons or axions. The coherent enhancement in this case can be interpreted either as the same bosonic $\sqrt{n+1}$ amplitude factor, in the quasiparticle picture, or, geometrically, as the result of the larger electric dipole moment of higher-$n$ transitions due to their larger radius.

\subsubsection{Coherent States}\label{subsub:CoherentMeasurement}

Since Dicke states are difficult to produce at large $N$, one can instead perform measurements with easily-prepared coherent states of qubits. This is the basis for typical atom interferometry within the standard quantum limit~\cite{InterferometerCoherenceReview}; in scattering contexts, it can be understood as the unsqueezed version of Refs.~\cite{SuperradiantDetection1, SuperradiantDetection2, SuperradiantDetection3}. The equatorial spin coherent state $\ketQ{+}^{\otimes N}$ is a binomial superposition of $O(\sqrt{N})$ Dicke levels, polarized along $x$ with $\expect{\mathbf{J}_x} = J = N/2$ and isotropic transverse fluctuations
\begin{equation}\label{eq:cohvar}
    \expect{\mathbf{J}_y^2} = \expect{\mathbf{J}_z^2} = \frac{J}{2} = \frac{N}{4},
    \qquad \sigma(\mathbf{J}_z) = \frac{\sqrt{N}}{2}.
\end{equation}
Because the state spreads over many Dicke (Fock) levels, a single bit flip (single added photon) perturbs it only at $O(1/\sqrt{N})$, so---unlike Dicke (Fock) states---coherent detector states are insensitive to one exchanged quantum, and the quantum Fisher information for a transverse rotation is only $F_Q = 4\expect{\mathbf{J}_z^2} = N$, the standard quantum limit. They can nonetheless register coherent interactions. A signal that flips spins in one direction (e.g. only down to up) shifts $\expect{\mathbf{J}_z}$ linearly, and thus becomes visible once the shift exceeds $\sigma(\mathbf{J}_z) \sim \sqrt{N}$, i.e. after $O(\sqrt{N})$ spin flips (photon additions); a signal of random sign instead diffuses $\mathbf{J}_z$, raising its variance by the number of flips, and so becomes visible only after $O(N)$ random spin flips (random photon additions or removals). 

The bosonic coherent state $\ketC{\alpha}$ behaves identically under the HPT: with $\expect{n} = |\alpha|^2$ and number spread $\sigma_n = |\alpha|$, two coherent states become distinguishable, $|\langle \alpha | \alpha + \delta \rangle| = e^{-|\delta|^2/2}$, only once $|\delta| \sim 1$---that is, once $O(|\alpha|)$ photons have been added in phase. Interactions that add and remove individual photons with equal probability (or, more generally, which lead only to unbiased diffusion in the bosonic Wigner plane) instead need $O(|\alpha|^2)$ photons to broaden the number distribution appreciably.

Since any laser constitutes a coherent state of photons, experiments employing such states are extremely widespread. We consider the PVLAS experiment~\cite{Ejlli:2020yhk} as an instructive example. In it, a linearly polarized laser passes through a cavity containing a magnetic field before being measured. A variety of effects---vacuum birefringence due to the Euler-Heisenberg Lagrangian, or photon-to-axion conversion through the inverse Primakoff effect, for example---can cause the presence of this magnetic field to change the polarization of the outgoing laser, whether by creating a non-zero ellipticity or by rotating the polarization angle through a small $\psi$. The number of photons scattered into the orthogonal polarization is then $N\sin^2\psi \approx N\psi^2$, linear in the integrated number of photons $N$; for a given minimum observable orthogonal polarization contribution this is a reach $\psi_{\rm min} \propto 1/\sqrt{N}$, the same coherent-state (shot-noise) scaling found above.

An alternative implementation of particle detection using bosonic coherent states is that of ion traps~\cite{Budker:2021quh}. These are loosely analogous to the electron cyclotrons discussed above, but now hold collections of ions in the ground state (or in a thermal state heavily concentrated near the ground state). The ground state is, however, a coherent state itself (and maps to the $\ketQ{0}^{\otimes N}$ coherent state via the HPT), so this can be understood as analogous to the above. Detection then occurs via observation of anomalous heating, in which the ions end up in states displaced from the ground state, but without relying on a particular magnitude or direction of that displacement. This is simply the small-occupation ($\hat{n} \to 0$) limit of the diffusive case above: starting from the vacuum $\ketC{0}$, any random displacement is in principle visible, but there is also no enhancement due to occupation number (although one can benefit from having many ions, i.e. many copies of the bosonic state; see Section~\ref{sec:NmScaling} below).

\subsubsection{Squeezed States}\label{subsub:SqueezedProtocols}

An intermediate approach between the Dicke and coherent cases above is to employ spin-squeezed states, and this has been done in a variety of contexts; see e.g. Refs.~\cite{Gross2010,Sewell2012,Hosten2016}. In these experiments, the reduced uncertainty of the squeezed state along one axis, relative to that of a coherent state, allows for improved sensitivity. This improvement is greatest when one can align the state's preferred direction to the (known) direction of a signal, but it can persist even if the signal has a random direction, because that direction will generically still have order-one overlap with the optimal one; this can be net-beneficial even if the state has lost sensitivity in other directions. Regardless, these squeezed spin states have immediate analogues in the boson literature, with squeezed photons states employed for various purposes in e.g. Refs.~\cite{Aasi2013,Wolfgramm:2010not}.

The HPT correspondence between squeezed states becomes somewhat more interesting, however, if one considers the dynamics of producing that squeezing. A standard approach to producing squeezing in qubit systems is one-axis twisting (OAT) \cite{Kitagawa1993,Gross2010,Hosten2016}, in which a Hamiltonian quadratic in $\mathbf{J}_z$ is applied to a coherent state in the $x$-$y$ plane. This results in, effectively, a larger frequency for Dicke state components at higher $M$, resulting in a ``stretching'' of the initially-round coherent state around the Bloch sphere; this produces squeezing. This process can immediately be mapped to the bosonic side, where the corresponding quadratic Hamiltonian contribution is a Kerr nonlinearity \cite{Kitagawa1986,Kirchmair2013}:
\begin{align}
    H \supseteq \omega \mathbf{J}_z + \chi \mathbf{J}_z^2 \HPTeq H \supseteq \left(\omega - N\chi\right) \hat{n} + \chi \hat{n}^2
\end{align}
up to an unphysical additive constant\footnote{In fact this holds anywhere on the Bloch sphere, not just at the equator, but the equator is the optimal place to perform OAT.}. The Kerr nonlinearity stretches a displaced bosonic coherent state around the origin, forming an elongated arc qualitatively resembling the lower-right panel of Fig.~\ref{fig:squeezed}. Unlike the genuinely number-squeezed state shown in that figure, however, the Kerr echo alone does not reduce the variance of $\hat{n}$, since $\hat{n}$ commutes with the Kerr Hamiltonian. This can be remedied using a subsequent rotation around the arc's center, just as a rotation must be performed at the end of OAT to align the produced state's squeezed direction with $\mathbf{J}_z$. Generating the quadrature-squeezed state in the upper-right of Fig.~\ref{fig:squeezed} requires a different procedure---parametric amplification---which we discuss below.

The squeezing potential of OAT is limited by over-twisting: as the stretched distribution bends around the Bloch sphere, its curvature eventually prevents further reduction of the uncertainty. (Longer application of OAT or the Kerr non-linearity can, however, be used to produce GHZ and cat states instead \cite{Molmer1999,Yurke1986,Kirchmair2013}.) This limitation can be remedied using an alternative technique: two-axis twisting (TAT), also referred to as two-axis counter-twisting (TACT) \cite{Kitagawa1993,Liu2011}. In TAT, one instead applies the Hamiltonian\footnote{Note that this conventional TAT Hamiltonian squeezes along a different direction (and thus its HPT map also squeezes along a different direction) than what is presented in Section~\ref{sub:SqueezedStates}. Both HPT maps are correct; the two are simply rotated relative to one another.}
\begin{align}
    H \supseteq \chi \left( \mathbf{J}_x^2 - \mathbf{J}_y^2 \right) \HPTnp H \supseteq \chi J \left( (a^\dagger)^2 + a^2 \right)
\end{align}
where we have immediately written down the bosonic analogue: the standard, single-mode squeezing Hamiltonian, often implemented using parametric amplification \cite{Wu1986}. (Here we omit $\omega$, assuming that either the detuning is taken to zero or that the squeezing Hamiltonian is applied in a co-rotating manner.) Note that the appropriate HPT map here is at the pole, not the equator; although the same parametric amplification Hamiltonian can be applied to displaced coherent states, this is not true of the TAT Hamiltonian. The advantage of TAT and parametric amplification is that they can generate arbitrarily large squeezing (up to finite-$N$ effects on the qubit side), and that they can do so more quickly, with squeezing growing exponentially in application time. While TAT is typically very difficult to implement in qubit contexts, parametric amplification is already used in dark matter detection in bosonic systems, most notably in the HAYSTAC experiment~\cite{Backes2021,Jewell2023}; see also Ref.~\cite{Malnou2019} for a more general discussion.

One additional approach to using squeezing in qubit-based measurements is noteworthy: squeeze-unsqueeze protocols \cite{Davis2016,Colombo2022,Hosten2016}. Squeeze-unsqueeze protocols assist in situations where one is limited by readout, rather than by the intrinsic uncertainty of the qubit state. In particular, in atomic contexts, it is often impractical to measure $\mathbf{J}_z$ with single-atom resolution, and thus squeezing a state's $\mathbf{J}_z$ uncertainty below the readout uncertainty no longer meaningfully improves precision. Squeeze-unsqueeze protocols remedy this by acting as an in-situ amplification, increasing both the change in the state and the corresponding intrinsic uncertainty by the same amount, rendering readout error less significant. It is implemented by applying an OAT squeezing period before the measurement time, and then applying it with the opposite sign after that measurement:
\begin{align}\label{eq:TwistUntwist}
    U = e^{-i\chi\tau \mathbf{J}_z^2} \, U_{\rm meas} \, e^{+i\chi\tau \mathbf{J}_z^2},
\end{align}
where $U$ is the full unitary evolution for the protocol and $U_{\rm meas}$ is the evolution during the measurement interval. In the absence of any non-trivial evolution during the measurement time, the two OAT intervals simply cancel out; a displacement along the squeezed direction, however, is amplified by precisely the squeezing factor, leading to a final state that is coherent but is displaced by a squeezing-enhanced angle.

The squeeze-unsqueeze protocol can immediately be translated to bosonic systems via the HPT: one applies Kerr nonlinearities of opposite signs before and after the measurement period. However, as in squeeze-unsqueeze, the final Kerr operation cannot increase the information encoded in the sensing state; its benefit is simply to make that information more accessible when one is limited by readout, which is less typical on the boson side of the map. In particular, since the Kerr nonlinearity commutes with $\hat{n}$, such a protocol provides no benefit when followed directly by number counting, which is widely available in bosonic systems \cite{Gleyzes2007,Dixit:2020ymh,Peil1999}. Although one might hope to still apply it in contexts where the number of bosons is too large to count individual photons (e.g. macroscopic cavity fields), photon loss severely restricts the advantage of single-mode squeezing over passive photon counting \cite{Shi:2024qhd}. We are therefore aware of no currently-existing systems in which Kerr-based squeeze-unsqueeze would be advantageous.

One can in principle also perform squeeze-unsqueeze-based measurements using TAT instead of OAT~\cite{Anders:2018ywx}, and a protocol closely related to this has recently been implemented~\cite{Gao:2025whg}. The HPT map of this is ${\rm SU}(1,1)$ interferometry, in which parametric amplifiers replace the beam splitters of a conventional interferometer \cite{Yurke:1986ipz,Caves:2019sxa}. In its balanced form, the protocol consists of squeezing, signal accumulation, and inverse squeezing: precisely the spin squeeze-unsqueeze sequence above but now using the bosonic analogue of TAT rather than OAT. This does not resolve the limitations above, however, and so we are similarly not aware of any direct applications of ${\rm SU}(1,1)$ interferometry to dark matter detection.

\subsubsection{GHZ States and Cat States}\label{subsub:GHZCatMeasurement}

Finally, several recent proposals describe the advantage of using GHZ states for detecting light dark matter~\cite{Chen:2023swh, Ito:2023zhp, Bodas:2025vff}. The GHZ state along $x$ of Eq.~\eqref{eq:GHZinFockBasis} is an equal superposition of the two $x$ poles, so $\expect{\mathbf{J}_x} = 0$ while $\expect{\mathbf{J}_x^2} = J^2 = N^2/4$ is quadratic in $N$ (the remaining $\expect{\mathbf{J}_y^2} = \expect{\mathbf{J}_z^2} = J/2$ are linear). This single large variance gives Heisenberg-scaling sensitivity to rotations about $x$, $F_Q = 4\expect{\mathbf{J}_x^2} = N^2$, matching the Dicke state up to an $O(1)$ factor. Like the Dicke state, it is also sensitive to a single spin flip: in the Fock basis of Eq.~\eqref{eq:GHZinFockBasis} it is supported only at even $J+M$, so a coherent bit flip changes $M$ by $1$ (while leaving $J$ unchanged) and thus yields a necessarily orthogonal, odd-$(J+M)$ state. GHZ states are also easily prepared with $O(N)$ gates, as described in Refs.~\cite{Chen:2023swh, Ito:2023zhp, Bodas:2025vff}. Alternatively one can simply measure the parity of the number of qubits in $\ketQ{1}$ before and after the integration time; notably there is not even a need to postselect on even parity, since the all-odd-$(J+M)$ state works essentially equally well. However a disadvantage of GHZ states---which we will remedy below---is that they cannot effectively distinguish a dark matter-induced excitation ($M \to M+1$) from a background decay ($M \to M-1$): both flip the parity, and because the state spreads over $\sim\sqrt{N} \gg 1$ Dicke levels the two outcomes are nearly identical. Writing $\ket{\psi_\pm} \equiv \mathbf{J}_\pm \GHZxN{} / \lVert \mathbf{J}_\pm \GHZxN{} \rVert$ for the normalized post-excitation and post-decay states, a straightforward calculation gives
\begin{equation}\label{eq:ghzoverlap}
    \big|\langle \psi_+ | \psi_- \rangle\big| = \frac{N-1}{N+1}
    \;\xrightarrow[N\to\infty]{}\; 1,
\end{equation}
so a parity measurement alone cannot tell signal from background.

Because of this difficulty, as well as the typically substantial decay rate of cavity photons, we are not aware of any precise bosonic analogues to this approach, i.e. of any experiments looking for dark matter using standard cat states~$\catN{2}$, although the strategy described in Refs.~\cite{Suri:2025hqm, Ma:2026ujz} can be understood as loosely analogous to doing so. An improved version of this approach has already been implemented, however: Ref.~\cite{Zheng:2025qgv} describes a dark photon search based on the four-legged cat state~$\catN{4}$. This performs similarly to the GHZ approach above, but with a crucial difference: since the $\catN{4}$ state is supported only at Fock numbers $n \equiv 0 \pmod{4}$, it can distinguish a dark matter-induced excitation, which raises $n \equiv 0 \pmod{4}$ to $n \equiv 1 \pmod 4$, from a cavity-loss decay, which lowers it to $n \equiv 3 \pmod 4$. Backgrounds must then arise from either three decays within one excitation time (taking $n \equiv 0 \to 3 \to 2 \to 1 \pmod{4}$, and so mimicking a signal), which is heavily suppressed as long as the decay timescale is longer than the integration time, or from a non-dark matter excitation, which is thermally suppressed whenever the photon energy exceeds the temperature.

These advantages of the $\catN{4}$ state over the $\catN{2}$ state motivate translating this back to the qubit side via the HPT, i.e. sensing using the GHZ-4 state. This is the subject of Section~\ref{sec:DetectionUsingGHZ4}.

\section{Dark Matter Detection Using the GHZ-4 State}\label{sec:DetectionUsingGHZ4}

As discussed in Section~\ref{sub:Z4States}, the GHZ-4 state carries a modulo-4 structure in the $\ketD{J,M}$ basis that is entirely analogous to the modulo-4 structure of the 4-legged cat state. It thus inherits the ability to distinguish decay errors (which reduce $M$ by 1) from signals that increase $M$, potentially offering a large improvement in background. This comes at only a small price in signal: Writing $\mathbf{J}_x^2 = (\mathbf{J}_+^2 + \mathbf{J}_-^2 + \mathbf{J}_+\mathbf{J}_- + \mathbf{J}_-\mathbf{J}_+)/4$, the off-diagonal $\mathbf{J}_\pm^2$ pieces connect $M \to M\pm 2$. For the GHZ-2 state these survive---its two $x$ poles give the full $\expect{\mathbf{J}_x^2} = J^2 = N^2/4$---but the $\mathbb{Z}_4$ projection retains only every fourth Dicke level, so they are projected away and only the diagonal part remains:
\begin{equation}
    \expect{\mathbf{J}_x^2} = \frac{J(J+1) - \expect{\mathbf{J}_z^2}}{2} \xrightarrow[N \gg 1]{}
    \frac{N^2+N}{8}. \label{eq:GHZ4Jx2}
\end{equation}
The quantum Fisher information for a rotation generated by $\mathbf{J}_x$ is therefore $F_Q = 4\expect{\mathbf{J}_x^2} = (N^2+N)/2$ for large $N$, approaching half the GHZ-2 value $F_Q = N^2$. This factor-of-2 price in signal can in many cases be far outweighed by the substantial background suppression, which we quantify below.

\begin{figure}[bt]
  \centering
  \includegraphics[width=0.6\textwidth]{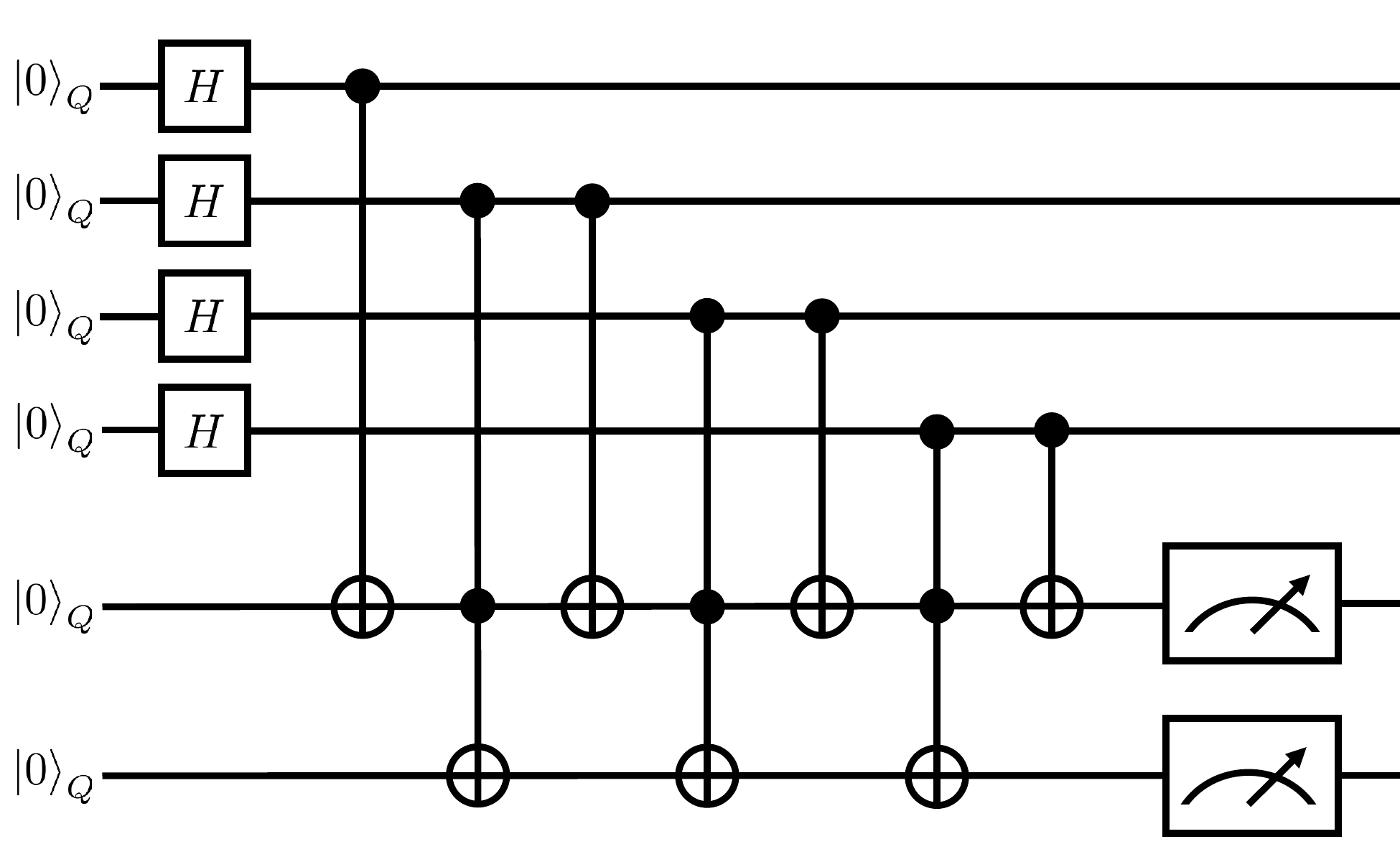}
  \caption{A simple circuit illustrating the preparation of the GHZ-4 state (and, omitting the Hadamard gates, the measurement process after integration). Here the first four qubits are those that will form the GHZ-4 state, while the last two are ancillae used to measure $\mathbf{J}_z$ modulo 4 of those four qubits; $J$ is fixed to $N/2$ in the absence of errors. A measurement of $J+M \equiv 0 \pmod{4}$ (e.g. the top ancilla in $\ketQ{0}$ and the bottom in $\ketQ{1}$, for four detector qubits), corresponds to the GHZ-4 state proper. However, as we note in the main text, the states corresponding to other definite values modulo 4 work just as well in the large-$N$ limit, potentially eliminating the need for any postselection. After an integration time, one simply measures $J_z \pmod{4}$ a second time, with an increase by one unit interpreted as a signal; decay noise will correspond instead to a decrease and can be distinguished for fewer than 3 decay errors.}
  \label{fig:GHZ4Circuit}
\end{figure}

\begin{figure}[tb]
  \centering
  \includegraphics[width=0.8\textwidth]{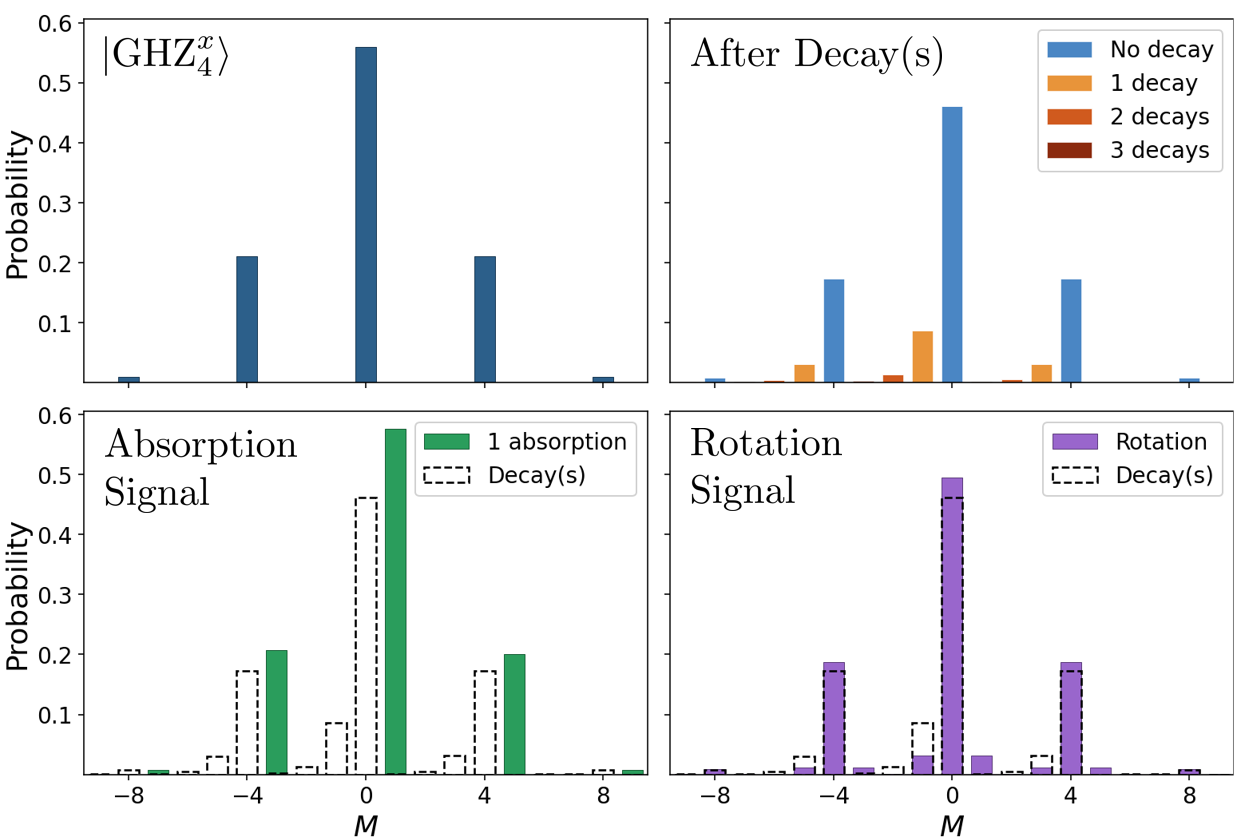}
  \caption{An illustration of the $M \pmod{4}$ structure (where $M$ is the value of the $\mathbf{J}_z$ operator, with the probability of each possible measurement outcome plotted) of the GHZ-4 state with $N=32$, as well as of the final states that may evolve from it.  (Note that, since $J=16$ here, $M$ and $M+J$ are equivalent modulo 4.) \textit{Top left:} The original $\GHZxN{4}$ state, whose support only on $M \equiv 0 \pmod{4}$ is apparent. We show only $|M| \leq 9$ since the GHZ-4 state's support at $|M|=12$ and $16$ is extremely small. \textit{Top right:} The distribution of $M$ values after 0, 1, 2, or 3 photon decays, starting from the $\GHZxN{4}$ state. Each decay lowers $M$ by 1. \textit{Bottom left:} The distribution of $M$ values after one photon is absorbed, e.g. due to dark matter. This increases $M$ by 1, thus giving a state with support only at $M \equiv 1 \pmod{4}$, which is distinguishable both from the initial state and from the results of 1 or 2 photons lost. A mistaken signal thus requires at least 3 losses, which is highly suppressed; the dotted envelope of the post-decay state (as in the upper-right plot) shows this. \textit{Bottom right:} The distribution of $M$ values after a small rotation, e.g. due to dark matter. This corresponds to evolution under a Hamiltonian $H \propto \mathbf{J}_y$, which gradually produces all possible $M \pmod{4}$ values, but changes $M$ only by $\pm 1$ at lowest order. A detection of $M \equiv 1 \pmod{4}$ after the integration time then offers a clean signal of dark matter since, as in the discrete absorption case, this cannot be confused with fewer than 3 photon losses. Note that the apparently much smaller signal in this case is an artifact of the lower-left plot being conditioned on one absorption; typically one expects zero dark matter absorptions during most integration times, whereas in the rotation case one sees consistent but small rotations.}
  \label{fig:GHZ4Support}
\end{figure}

We illustrate a circuit that can be used to produce the GHZ-4 state in Figure~\ref{fig:GHZ4Circuit}, which is nothing more than a preparation of the coherent-along-$x$ state $\ketQ{+}^{\otimes N}$ followed by a mod-4 measurement of $J_z$. In the event that the measurement projects the qubits into $J+M \equiv 0 \pmod{4}$, this immediately gives the $\GHZxN{4}$ state, as shown in Eq.~\eqref{eq:GHZ4inFockBasis}. Any other measurement will give a different state---one supported only at $J+M \equiv 1$, $2$, or $3 \pmod{4}$---but it is easily verified that these work just as well for $N \gg 1$. 

Following this initial ``preparation'' measurement, the state can then be left to evolve freely for some time. A decay error during this integration time corresponds to a decrease in $M$ by $1$; thus an initially $M \equiv 0 \pmod{4}$ state will become supported only at $M \equiv 3 \pmod{4}$ (although it will no longer be fully permutation-symmetric, for a decay of a specific qubit, and thus cannot be written in the $\ketD{J=N/2,M}$ basis). Conversely, a dark matter excitation would take $M \equiv 0 \pmod{4}$ to $1 \pmod{4}$. Both possibilities are illustrated in Figure~\ref{fig:GHZ4Support}. Notably this is true regardless of whether the excitation is coherent over all of the qubits, although, again, an incoherent excitation (e.g. a hard scattering event) will remove the state from the Dicke manifold. Regardless, the same mod-4 measurement shown in Figure~\ref{fig:GHZ4Circuit}, only without the Hadamard gates, will then reveal what has occurred, with no possibility of a false signal unless at least 3 decays have occurred.

This discussion generalizes straightforwardly to GHZ-$k$ states for generic $k$, with only marginally more complicated circuits for $k$ that is not a power of $2$. The analysis above shows that all $\GHZxN{k}$ with $k>2$ share the same signal sensitivity, as the $\mathbb{Z}_k$ projection removes the $M\to M\pm 2$ coherences for every $k \geq 3$, so $F_Q = (N^2+N)/2$ independently of $k$ (for $k \geq 3$), in the $N \gg k^2$ limit. The factor of two relative to GHZ-$2$ is therefore paid only once, at $k = 2 \to 3$; larger $k$ buys distinguishability from up to $k-2$ decay errors at no further cost in signal, except due to the gate error effects discussed below. (There is an additional factor-of-2 cost in signal when observing only the $\Delta M = +1$ part of a rotation, but again this does not grow with large $N$ or $k$.)

Quantifying the resulting background suppression is straightforward: Let $\lambda$ be the mean number of decays during one integration window, which we assume to be $\ll 1$. Decays are independent, so the probability of the $k-1$ successive decays needed to mimic a signal (taking $M \equiv 0 \to k-1 \to \cdots \to 1 \pmod{k}$, for example) is Poisson-suppressed as $\lambda^{k-1}/(k-1)!$. The GHZ-2 state is faked by a single decay, with false-signal probability $\sim\lambda$; the GHZ-4 state requires three, suppressing the decay background to $\sim\lambda^3/6$. There is, however, an additional irreducible floor of thermal excitations: by detailed balance, these occur with mean number $\sim \lambda\,e^{-\omega/T}$ per observation period, with $\omega$ the qubit splitting and $T$ the temperature. 

An additional source of false signals is imperfect gates. Since preparation and readout together use $\approx N(1+2\log_2(k))$ gates, a per-gate error rate $\epsilon$ will contribute $\sim \epsilon N (1+\log_2 k)$ expected false signals, omitting any gate-type-dependence of $\epsilon$ for simplicity. The total false-signal probability using a GHZ-$k$ state is thus
\begin{equation}\label{eq:ghzbkg}
    P_{\rm false} \sim \underbrace{\frac{\lambda^{\,k-1}}{(k-1)!}}_{\rm decays}
    + \underbrace{\lambda\,e^{-\omega/T}}_{\rm thermal} + \underbrace{\epsilon N (1+\log_2 k)}_{\rm imperfect~gates},
\end{equation}
and the role of increasing $k$ is simply to push the first term below the second. There is negligible benefit to increasing $k$ once the thermal term dominates the false-signal probability, so even in the well-cooled regime $\omega \gg T$ a modest $k$ is likely to be sufficient; larger $k$ are likely to instead be harmful as they increase the gate error contribution. We leave a detailed analysis of this technique's performance using realistic imperfect gates, as well as using alternative preparation and readout sequences, to future work.

Notably, once $k > N$, the $\GHZxN{k}$ state is simply the $M=-J$ Dicke state; the procedure above without postselection would likewise give a Dicke state at some random $M$, typically with $|M| \lesssim \sqrt{N}$ since this is where the initial coherent state is primarily supported. The advantage of $\GHZxN{k}$ states for smaller $k$ is thus simply ease of preparation: although a sufficiently good quantum computer could prepare the Dicke state, producing a low-$k$ GHZ-$k$ state may be more feasible for near-term devices, and such states will offer essentially the entire benefit of the ideal Dicke state once $P_{\rm false}$ is dominated by thermal excitations.

\section{Sensitivity Scalings with Sensor Count and Excitation Number}\label{sec:NmScaling}

Although the HPT relations above ensure that bosonic and qubit systems are analogous up to constant factors in various limits, the constant factors in question can be parametrically large, being $O(\sqrt{N})$ unless $N-\hat{n} \ll N$ (see Eqs.~\eqref{eq:HPTGeneral1}--\eqref{eq:HPTGeneral3}). This will lead to e.g. factor-of-$N$ differences in the optimal coherent enhancement of quantum Fisher information (QFI) or cross-section in Section~\ref{sec:examples}, with bosonic systems enhanced by at most $O(N)$ while qubits can experience $O(N^2)$. Here we offer two perspectives on understanding this difference.

From one perspective, the factor of $O(\sqrt{N})$ advantage of qubits can be understood as the product of a conserved number: In a system of $N$ qubits, any process $\mathbf{J}_-$ that increases the number of qubits in $\ketQ{0}$ by one necessarily decreases the number of qubits in $\ketQ{1}$ by one. Thus, instead of associating $\mathbf{J}_-$ with $ \sqrt{N-\hat{n}}\, a$, we could associate it with $a b^\dagger$, where $b$ is the lowering operator of a bosonic mode containing $N - a^\dagger a$ bosons, such that the total number of bosons $a^\dagger a + b^\dagger b$ is exactly fixed to $N$. Such a number-conserving pair of bosonic modes is then \textit{exactly} equivalent to the $N$-qubit system, with one mode reflecting the number of qubits in $\ketQ{1}$ and the other the number in $\ketQ{0}$. This produces an exact HPT-like map, but at the price of introducing a second bosonic mode, which is typically unphysical; the $O(\sqrt{N})$ factors necessary on the bosonic side of the usual HPT is then the price of not having this additional mode.

An alternative perspective, however, can be obtained by considering systems with multiple bosonic modes: Although our discussion so far has focused on comparing a collection of qubits to a single bosonic mode via the HPT, one could instead use $m$ parallel bosonic modes, with $O(n)$ total excitations between them. (The most common form of this is a laser interferometer, whose two arms can be understood as two modes in this language, but a variety of examples exist \cite{Budker:2021quh, Freiman:2025tse, LIGOO4Detector:2023wmz, Mckenzie:2002ir, Guo:2019zef, Xia:2019jil}.) Sensitivity to wave-like dark matter can be parametrically enhanced by $m$; a collection of $m$ cavities with $n$ excitations spread symmetrically between them (i.e. the state $\propto (\sum_{k=1}^m a_k^\dagger)^n$, with $a_k^\dagger$ the raising operator for the $k$'th cavity) has a QFI of $F_Q = 4m(2n+1) \sim 4mn$ with respect to the operator $\sum_{k=1}^m \left(a_k + a_k^\dagger\right)$, i.e. the generator of collective displacements in the $p$ direction. This leads to the following schematic form for the optimal QFI enhancement of both qubit- and boson-based detectors, which we prove below:
\begin{equation}
    F_Q^{\rm opt} \sim F_Q^{(0)} \times \left(\mbox{number of sensors}\right)
    \times \left(\mbox{total occupation number}+1\right)
\end{equation}
where $F_Q^{(0)}$ is the QFI one would get with a single and unoccupied sensing component. In qubit systems, both the number of sensors and the maximum occupation number scale with $N$, resulting in the $O(N^2)$ scaling above. These two quantities are independent in bosonic systems, however, and thus their corresponding scaling is instead $O(nm)$.

\subsection{The Hamiltonian-Not-in-Lindblad-Span Condition for Qubits and Bosons}\label{sub:HNLS}

In fact, the analogy between optimal sensitivity scaling in qubit and boson systems holds not only in the optimal, noise-free case, but also in the presence of noise. A general framework for determining these optimal scalings is given by the Hamiltonian-Not-in-Lindblad-Span (HNLS) condition~\cite{Zhou2018HNLS,DemkowiczDobrzanski2017HNLS}. We briefly summarize its qubit formulation here before presenting a bosonic analogue.

Consider a system of $N$ qubits, used to detect a signal Hamiltonian $H_s = gG = g\sum_{k=1}^N G_k$ where each $G_k$ acts only on the $k$'th qubit. The system's noise is described by a collection of jump operators $\{L_{k,\lambda}\}_{k,\lambda}$, with $L_{k,\lambda}$ acting only on the $k$'th qubit for all $\lambda$. For simplicity we assume that the signal and noise are the same for all $k$ and time-independent. Then let
\begin{align}
    \mathcal{S} = {\rm span}\left( I, L_{k,\lambda}, L_{k,\lambda}^\dagger, L_{k,\lambda}^\dagger L_{\ell,\kappa} \right)
\end{align}
be the Lindblad span. The HNLS theorem asserts that $g$ can be estimated with Heisenberg scaling---i.e. $(\Delta g)^{-2} = O(N^2 T^2)$, where $T$ is the observation time---if and only if $G \notin \mathcal{S}$. For $G \in \mathcal{S}$, one can achieve only the Standard Quantum Limit, $(\Delta g)^{-2} = O(N T)$. Although quadratic scaling may be possible in the $G \in \mathcal{S}$ case for sufficiently short $T$ (essentially because the effects of noise are still negligible), only linear scaling in $N$ and $T$ is possible at large times; see the more precise bosonic discussion below. 

HNLS is often presented without assuming that signal and noise act on individual qubits independently, as we did above. In this case, however, only the statements about scaling with observation time hold; multi-qubit operators can change scaling with $N$ for both $G \in \mathcal{S}$ and $G \notin \mathcal{S}$.

The HNLS theorem also applies to bosonic systems, up to technical caveats related to their infinite-dimensional Hilbert spaces~\cite{Zhou2018HNLS,WanLasenby2022}. However, the factors of $N$ above do not immediately translate to specifying scaling with $m$ and/or $n$ for bosonic modes, although related bounds incorporating both time and energy in bosonic settings have been studied in Ref.~\cite{Gorecki2025}. Inspired by the HPT, we therefore derive the joint scaling with mode number $m$, mean total occupation $n$, and observation time $T$ for linear signal and noise operators, which will parallel the joint $N$ and $T$ scaling for qubit systems discussed above.

Consider $m$ parallel bosonic modes\footnote{Formally our proof also requires a maximum occupation number for each mode to ensure terms remain finite, but this can be made arbitrarily large; our results are independent of the cutoff.}, with $n = \overline{\expect{\sum_k \hat{n}_k}}$ expected total bosonic excitations between them, averaged over the observation time. We restrict to signals linear in the raising and lowering operators; up to a basis change, all such signals are simply $H = gG = g\sum_{k=1}^m x_k$, with $x_k$ the $x$ operator on the $k$'th mode. This restriction is necessary for the same reason as the individual-qubit-signal/noise assumption above: without it, one can straightforwardly obtain signals that scale as higher powers of $n$ simply by coupling to high powers of the mode occupation. We likewise assume linear noise: $\{L_{k,\lambda}\}_{k,\lambda}$ where $L_{k,\lambda} = \sqrt{\kappa} \left( u_\lambda x_k + v_\lambda p_k \right)$, with $p_k$ the $p$ operator on the $k$'th mode, $\kappa$ an overall noise rate, and $u_\lambda$, $v_\lambda$ some complex constants satisfying $|u_\lambda|^2 + |v_\lambda|^2 \leq 1$ which set the relative direction of the noise relative to the signal (and for multiple noise types, set their relative rates). We take $\kappa$, $u_\lambda$, and $v_\lambda$ to be $k$-independent for simplicity.

Then define the Lindblad span as before:
\begin{align}
    \mathcal{S} = {\rm span}\left( I, L_{k,\lambda}, L_{k,\lambda}^\dagger, L_{k,\lambda}^\dagger L_{\ell,\kappa} \right).
\end{align}
This gives a ``bosonic linear HNLS'' theorem, which we prove in Appendix~\ref{app:BosonicHNLS}:
\begin{align}
    \frac{1}{\left( \Delta g\right)^2} =
    \begin{cases}
        O\left( m(n+1)T^2 \right) & G \notin \mathcal{S},\\[4pt]
        O\left( m \min\left( (n+1)T^2, \dfrac{T}{\kappa} \right) \right) & G \in \mathcal{S}.
    \end{cases} \label{eq:BosonicLinearHNLS}
\end{align}
Moreover these parametric scalings are always achievable (under their respective conditions) for $n \geq 1$.

The $G \notin \mathcal{S}$ result of Eq.~\eqref{eq:BosonicLinearHNLS} is a more precise version of our ``optimal QFI'' statement above, producing a clean $N^2 \HPT mn$ correspondence. The $G \in \mathcal{S}$ result, conversely, establishes the same with-noise behavior known in the qubit case: although one can still scale with number of parallel sensors once the signal is within the noise, it is no longer possible to win parametrically by preparing highly-occupied states, whether those are entangled states of $N$ qubits or excited states of bosonic modes.

This bosonic linear HNLS result is practically somewhat trivial, since essentially all bosonic systems experience decays, $L_{k,{\rm decay}} \propto a_k$, and thus displacements will always be within $\mathcal{S}$ unless decays are so rare that one can omit them entirely. This is not necessarily the case for higher-order operators, however: two-photon signals such as 
\begin{align}
    H_s = g\sum_{k=1}^m \left(a_k^\dagger\right)^2 + a_k^2
\end{align}
are no longer within the span of decays alone, and attain $O((n^2 + m(n+1))T^2)$ scaling for a sensor state $(A^\dagger)^{n/2}\ket{0}^{\otimes m}$ where $A^\dagger = \sum_k (a_k^\dagger)^2$. A straightforward modification of the argument in Appendix~\ref{app:BosonicHNLS} can be used to, for example, extend our HNLS dichotomy to signal operators at second (or higher) order in $a_k$ and $a_k^\dagger$ as well (although with additional assumptions), but we do not present this in detail since the scaling with $m$ and $n$ becomes more involved and is not as cleanly analogous to qubit behavior.

\section{Summary of the Analogy}
\label{sec:summary}

In this work, we have described the Holstein-Primakoff transformation and explained how it connects a number of existing experimental techniques applicable to dark matter detection. The HPT connects the space, operators, and states of a collection of qubits to those of a (truncated) bosonic mode. The operator correspondence is given by Eqs.~\eqref{eq:HPTGeneral1}-\eqref{eq:HPTGeneral3}, with two simple limits presented in Section~\ref{sub:HPTLimits}; the HPT connections between states are summarized in Table~\ref{tab:StateCorrespondence}.

Although most typically presented in its polar limit, where the HPT corresponds to the geometric intuition of the Bloch sphere being locally flat---like the bosonic $(x,p)$ plane---the HPT is also conveniently approximated near any fixed ``latitude'' (i.e. excited state count), where it corresponds to an annulus of approximately-fixed occupation number in the bosonic mode. The latter is particularly useful in quantum sensing contexts, as it establishes a map between sensitive entangled states, for example between Dicke and Fock states, and between GHZ and cat states.

Our purpose in presenting the HPT, however, is not merely to connect known results: the HPT can additionally reveal novel sensing strategies when only one side of a correspondence has previously been identified. We illustrate this with the GHZ-$k$ states (for $k>2$), which are straightforward HPT analogues of the bosonic $k$-legged cat states. Although the advantage of $k$-legged cat states for dark matter detection---namely that it allows one to distinguish a dark matter-induced excitation from the dominant decay background---is known, this advantage has not previously been passed to the qubit side of the HPT. The GHZ-$k$ state does exactly this and, as we show, it is both straightforwardly prepared and has the same advantage as the $k$-legged cat for dark matter detection.

States and protocols are not the only things that can be mapped through the HPT however: even more general results can be analogous between the two sides. The number-scaling result of the Hamiltonian-Not-in-Lindblad-Span theorem, for example, establishes when optimal measurement and control of a system of $N$ qubits allows for uncertainties that decrease as $N^{-1/2}$ versus $N^{-1}$; in this work we showed that this corresponds to uncertainties scaling as $m^{-1/2}$ and $(mn)^{-1/2}$, respectively, for $m$ bosonic modes containing a total of $n$ excitations.

These are certainly not the only novel results that can be obtained with the HPT, however, and we present them here solely as motivating examples. We leave further exploration of HPT-inspired states, sensing strategies, and results to future work.

\begin{center}
\begin{table}
    \renewcommand{\arraystretch}{1.6}
    \begin{tabular}{|c|c|c|}
    \hline
    \multicolumn{3}{|c|}{\textbf{HPT Between States}} \\
    \hline
    \textbf{Qubits (symmetric, $J = N/2$)} & Section & \textbf{Bosons (single mode)} \\
    \hline
    All spins down, $\ketQ{0}^{\otimes N} = \ketD{J,-J}$ & \ref{sub:VacuumHPT} & The vacuum, $\ketF{0}$ \\
    \hline
    Dicke states, $\ketD{J,M}$ & \ref{sub:DickeFockHPT} & Fock states, $\ketF{J+M}$ \\
    \hline
    Coherent states, $(\sin\frac{\theta}{2}\ketQ{0} + e^{-i\phi}\cos\frac{\theta}{2}\ketQ{1})^{\otimes N} = \ketD{\theta,\phi}$ & \ref{sub:CoherentStates} & Coherent states, $\ketC{\sqrt{N}e^{-i\phi}\cos\frac{\theta}{2}}$ \\
    \hline
    The coherent state along $+x$, $\ketQ{+}^{\otimes N} = \ketD{\theta=\frac{\pi}{2}, \phi=0}$ & \ref{sub:CoherentStates} & The coherent state $\ketC{\alpha = \sqrt{J}}$ \\
    \hline
    GHZ states (along $x$), $\ketQ{+}^{\otimes N}+\ketQ{-}^{\otimes N}$ & \ref{sub:GHZCatHPT} & Cat states, $\ketC{\alpha} + \ketC{-\alpha}$ \\
    \hline
    GHZ-$k$ states, $\Pi_{J+M \equiv 0 \pmod{k}} \ketQ{+}^{\otimes N}$ & \ref{sub:Z4States} & $k$-legged cat states, $\sum_{\ell=0}^{k-1} \ketC{\alpha = e^{2\pi i\ell/k}\sqrt{J}}$ \\
    \hline
    \end{tabular}
    \caption{The HPT correspondence between various states of spin/qubits and of a bosonic mode, as well as the sections of this work in which we discuss them. This table omits squeezed states due to an ambiguity in their HPT correspondence; see Sections~\ref{sub:SqueezedStates} and~\ref{subsub:SqueezedProtocols}}
    \label{tab:StateCorrespondence}
\end{table}
\end{center}

\acknowledgments

The authors thanks Yao Lu, Hajime Fukuda, Ryan Plestid, Clara Murgui, Leonardo Badurina, Sydney Timmerman, Savas Dimpoulos, and Marios Galanis for insightful conversations. 
This work was supported by the
Office of High Energy Physics of the U.S. Department
of Energy QuantISED program. RH is also partially
supported by the Department of Energy, Office of Science, National Quantum Information Science Research
Centers, Superconducting Quantum Materials and Systems Center (SQMS) under contract number DE-AC02-
07CH11359. This manuscript has been authored by
Fermi Forward Discovery Group, LLC under Contract
No. 89243024CSC000002 with the U.S. Department of
Energy, Office of Science, Office of High Energy Physics.
This work was performed in part at the Aspen Center for Physics, which is supported by National Science Foundation grant PHY-2210452. 

Several versions of GPT-5.x/6.x, Opus-5.x, and Fable-5.x were used to assist in plot generation, performing preliminary calculations, and finding references. However all final calculations were performed by the authors, and all text was written by the authors, up to LLM-assisted proofreading.

\bibliographystyle{apsrev4-2}
\bibliography{main}

\appendix

\section{Wigner Functions}
\label{app:WignerFunctions}

Throughout the main text we visualize states using Wigner quasi-probability functions: the standard bosonic Wigner function on the phase-space plane, and its spin counterpart on the Bloch sphere. Both can be described in a common language: In both cases, one first chooses some operator at a reference point, and transports it through phase space. The Wigner function at a point is then the expectation value of the corresponding transported operator. For a bosonic mode the relevant transformations are displacements on the plane, while for a spin they are rotations on the sphere. We define both constructions here and fix our conventions, since factors of $2$ and $\pi$ vary between references and software packages. All plots are generated with QuTiP~\cite{QuTiP}.

\subsection{The bosonic Wigner function}
\label{subapp:BosonWigner}

The starting point for the bosonic Wigner distribution is the bosonic number parity operator
\begin{equation}
    \Pi=(-1)^{a^\dagger a},
\end{equation}
which reflects phase space through the origin,
\begin{equation}
    \Pi x\Pi = -x, \qquad \Pi p\Pi = -p.
\end{equation}
Displacing parity moves the center of this reflection to the point $\alpha = (x + ip)/\sqrt{2}$:
\begin{equation}
    \Pi(\alpha) = D(\alpha)\Pi D^\dagger(\alpha), \qquad D(\alpha) = e^{\alpha a^\dagger - \alpha^* a}.
\end{equation}
The Wigner function is the expectation value of this displaced parity,
\begin{align}
\begin{split}
    W(x,p) &= \frac{1}{\pi} \Tr\left[ \rho \Pi(\alpha) \right] \\
    &= \frac{1}{2\pi} \int_{-\infty}^{\infty} dy \, e^{-ipy} \left\langle x + \frac{y}{2} \middle| \rho \middle| x - \frac{y}{2} \right\rangle . \label{eq:bosonwigner}
\end{split}
\end{align}
The second line is the usual position-space definition, which shows that $W(x,p)$ depends on the coherence between pairs of positions centered at $x$. The first line offers an alternative definition that will be most natural for this discussion: displace the state by $-\alpha$ and subtract the odd photon-number probability from the even one.

Since parity is Hermitian with eigenvalues $\pm1$, $W$ is real and bounded, $|W|\leq1/\pi$. It is normalized as
\begin{equation}
    \iint dx\,dp\, W(x,p) = 1,
\end{equation}
and its marginals are the quadrature probability distributions,
\begin{equation}
    \int_{-\infty}^{\infty} dp \, W(x,p) = \left\langle x \middle| \rho \middle| x \right\rangle, \qquad \int_{-\infty}^{\infty} dx \, W(x,p) = \left\langle p \middle| \rho \middle| p \right\rangle.
\end{equation}
$W(x,p)$ being negative anywhere implies that a state is non-classical.

In this convention, the vacuum's Wigner function is the symmetric Gaussian $W(x,p) = e^{-(x^2+p^2)}/\pi$, while that of a coherent state $\ketC{\alpha}$ is the same Gaussian displaced to
\begin{equation}
    (x,p) = \left( \sqrt{2} \; {\rm Re}(\alpha), \, \sqrt{2} \; {\rm Im}(\alpha) \right).
\end{equation}
The factors of $\sqrt{2}$ are a product of our choice of conventions, which matches the $g=\sqrt{2}$ default scaling of QuTiP's \texttt{wigner}; matching its $g=2$ scaling would eliminate these factors of $\sqrt{2}$ at the cost of introducing constants elsewhere in this discussion.

\subsection{The spin Wigner function on the Bloch sphere}
\label{subapp:SpinWigner}

Generating a sensible Wigner function for qubits is somewhat more involved. We begin by decomposing the density matrix $\rho$ of a given symmetric (i.e. $J = N/2$) state as
\begin{align}
    \rho = \sum_{k=0}^{N}\sum_{q=-k}^{k} \rho_{kq} T_{kq}
\end{align}
where $T_{kq}$ is a spherical tensor operator (normalized as $\Tr[T_{kq}^\dagger T_{k'q'}] = \delta_{kk'}\delta_{qq'}$) and
\begin{align}
    \rho_{kq} = \Tr\left[ \rho T_{kq}^\dagger \right].
\end{align}
The spherical tensors transform under rotations in the same way as the spherical harmonics $Y_{kq}$~\cite{SphericalTensorHarmonicBook}, which we use to define the spin Wigner function, following Refs.~\cite{Agarwal1981,Dowling1994}:
\begin{align}
    W(\theta,\phi) = \sum_{k=0}^{N} \sum_{q=-k}^{k} \rho_{kq} Y_{kq}(\theta,\phi). \label{eq:SpinWigner}
\end{align}
We use the spherical tensor and spherical harmonic conventions implemented in QuTiP's \texttt{spin\_wigner}.

The spin Wigner function is normalized by
\begin{equation}
    \int d\Omega \, W(\theta, \phi) = \sqrt{\frac{4\pi}{N+1}}.
\end{equation}
However, while the spin Wigner function is real, it does not need to be positive even for classical states, unlike the bosonic Wigner function. Moreover there is no straightforward means of recovering spin-projection probabilities by integrating over one coordinate, as there was for $W(x,p)$. Nonetheless the spin Wigner function does approach the HPT transform of the bosonic Wigner function in an appropriate limit, as we show below.

We can rewrite our definition of $W(\theta,\phi)$ as
\begin{align}
    W(\theta,\phi) = \Tr\left[ \rho \Delta(\theta,\phi) \right]
\end{align}
where
\begin{align}
    \Delta(\theta,\phi) = \sum_{k=0}^{N} \sum_{q=-k}^{k} Y_{kq}(\theta,\phi) T_{kq}^\dagger
\end{align}
is the ``Stratonovich-Weyl kernel'' (see e.g. Refs.~\cite{Agarwal1981, Amiet:2000rju}). 

Using the south pole as our reference point ($\ketD{J,-J} \HPT \ketF{0}$), as in the rest of this work, and noting that $Y_{kq}(\pi,\phi) = \delta_{q0} (-1)^k \sqrt{(2k+1)/(4\pi)}$, we have
\begin{align}
    \Delta_{\rm SP} = \sum_{k=0}^{N} (-1)^k \sqrt{\frac{2k+1}{4\pi}} T_{k0}
\end{align}
at the south pole. Rotational covariance then gives us
\begin{align}
    W(\theta,\phi) = \Tr\left[ \rho R(\theta,\phi) \Delta_{\rm SP} R^\dagger(\theta,\phi) \right], \label{eq:SpinWignerKernelRotation}
\end{align}
where
\begin{align}
    R(\theta,\phi) = e^{ i \left(\pi-\theta\right) \left( -\mathbf{J}_x \sin\phi + \mathbf{J}_y \cos\phi \right) }
\end{align}
is a rotation of the south pole to $(\theta,\phi)$. Eq.~\eqref{eq:SpinWignerKernelRotation} is thus the qubit analogue of Eq.~\eqref{eq:bosonwigner}, defining the spin Wigner function as the expectation of a rotated Stratonovich-Weyl kernel rather than a displaced parity operator.

Ref.~\cite{Amiet:2000rju} shows\footnote{Although note that it follows north pole rather than south pole conventions.} that 
\begin{align}
\begin{split}
    \left\langle J,-J+n \middle| \Delta_{\rm SP} \middle| J,-J+n \right\rangle &= \sqrt{\frac{N+1}{4\pi}} \sum_{k=0}^{N} \frac{2k+1}{N+1} \left\langle J, J-n; k, 0 \middle| J, J-n \right\rangle \\
    &= (-1)^n \sqrt{\frac{N+1}{4\pi}} \sum_{k=0}^{N} \sqrt{\frac{2k+1}{N+1}} \left\langle J, J-n; J, n-J \middle| k, 0 \right\rangle,
\end{split}
\end{align}
where the right-hand sides are sums of Clebsch-Gordan coefficients. It then evaluates the limit of that sum to
\begin{align}
    \lim_{J \to \infty} \sum_{k=0}^{N} \sqrt{\frac{2k+1}{N+1}} \left\langle J, J-n; J, n-J \middle| k, 0 \right\rangle = 2,
\end{align}
and thus we have
\begin{align}
    \lim_{J \to \infty} \sqrt{\frac{\pi}{N+1}} \left\langle J,-J+n \middle| \Delta_{\rm SP} \middle| J,-J+n \right\rangle = (-1)^n.
\end{align}

In the high-$N$, south pole limit ($J \to \infty$ at fixed $n$), where $\ketD{J,-J+n} \HPT \ketF{n}$ and displacements are the HPT maps of rotations, we therefore have
\begin{align}
    \Delta_{\rm SP} \HPTnp \sqrt{\frac{N+1}{\pi}} (-1)^{a^\dagger a}.
\end{align}
The Stratonovich-Weyl kernel is thus HPT-analogous at the south pole to the parity operator used to define the bosonic Wigner function, albeit only in the large-$N$ limit, and the resulting Wigner functions are also HPT-analogous in this limit:
\begin{align}
    W(\theta,\phi) \HPTnp \sqrt{\pi(N+1)} \, W(x,p), 
\end{align}
with a factor of $\pi$ appearing from the normalization of Eq.~\eqref{eq:bosonwigner}.

\section{A Proof of HNLS for Bosonic Modes}\label{app:BosonicHNLS}

In this appendix we prove our ``bosonic linear HNLS'' result, Eq.~\eqref{eq:BosonicLinearHNLS}. This consists of, effectively, four different statements: that the scalings in both cases are maximal (i.e. that one cannot scale faster with $m$, $n$, or $T$), and that the scalings are achievable (i.e. that it is possible to scale that fast). The first two subsections below prove the former: we first show that one can never exceed $O(m(n+1)T^2)$, and then that one is limited to $O( m \min( (n+1)T^2, T/\kappa ))$ when $G \in \mathcal{S}$. The last two subsections provide explicit examples satisfying these regimes for $n \geq 1$, first when HNLS holds and then when it does not. Our proof is based heavily on those in Refs.~\cite{Zhou2018HNLS, DemkowiczDobrzanski2017HNLS}, so we focus only on what changes in bosonic settings.

\subsection{The Upper Bound When HNLS is Satisfied}

In the absence of noise or control operations, this universal upper bound would be trivial: since the signal $G = \sum_k x_k$ acts only on the collective mode $b = m^{-1/2}\sum_k a_k$, with occupation $\expect{b^\dagger b} \leq n$, we have
\begin{align}
    \Var(G) \leq m\expect{x_b^2} \leq m\expect{x_b^2 + p_b^2} \leq m(2n+1),
\end{align}
where $x_b$ and $p_b$ are the $x$ and $p$ operators of the collective mode. The Quantum Fisher information (QFI), $F_Q$, is upper bounded by $4\Var(G)T^2$, and thus the Cram{\'e}r-Rao bound would imply our upper bound \cite{QFIReview}.

To accommodate a measurement with arbitrary noise and control, consider that measurement's purification~\cite{Kolodynski:2013lka}, i.e. the unitary evolution of some large system containing our sensors as well as their environment and the measurement registers. The true measurement's QFI is upper bounded by the purified QFI, which in turn is limited by the triangle inequality on the square root of the QFI, integrated over the observation time. We thus have
\begin{align}
    F_Q \leq 4m(2n+1)T^2 \sim m(n+1)T^2,
\end{align}
and therefore an upper bound, independent of whether $G \in \mathcal{S}$, of
\begin{align}
    \frac{1}{\left( \Delta g\right)^2} = O( m(n+1)T^2 ).
\end{align}

\subsection{The Upper Bound When HNLS is Not Satisfied}

To show the stronger upper bound when $G \in \mathcal{S}$, we closely follow the method and notation of Ref.~\cite{Zhou2018HNLS}. Since the $L_{k,\lambda}$ are linear in $a_k$ and $a_k^\dagger$, we can write
\begin{align}
    x_k = \sum_\lambda c_\lambda^* \frac{L_{k,\lambda}}{\sqrt{\kappa}} + c_\lambda \frac{L_{k,\lambda}^\dagger}{\sqrt{\kappa}}
\end{align}
for some coefficients $\{c_{\lambda}\}$. Note that the $c_\lambda$ are $k$-independent (and thus $m$-independent) because we assumed that $L_{k,\lambda}$ has the same direction for all $k$. Now let $q_{k,\lambda} = c_\lambda/\sqrt{\kappa}$, so
\begin{align}
    G = \sum_{k,\lambda} q_{k,\lambda}^* L_{k,\lambda} + q_{k,\lambda} L_{k,\lambda}^\dagger.
\end{align}
Then our infinitesimal channel Kraus operators~\cite{Kolodynski:2013lka} can be written as
\begin{subequations}
\begin{align}
    K_0 &= I - \left( igG + \frac{1}{2}\sum_{k,\lambda} L_{k,\lambda}^\dagger L_{k,\lambda} \right) dt + O\left(dt^2\right) \\
    K_{k,\lambda} &= L_{k,\lambda}\sqrt{dt} + O\left(dt^{3/2}\right).
\end{align}
\end{subequations}
Kraus operators are gauge-dependent. Let $h$ be a Hermitian matrix with $h_{0,(k,\lambda)} = -q_{k,\lambda}^*\sqrt{dt}$ with all other entries (except $h_{(k,\lambda),0}$) zero, and define $\widetilde{K} = e^{-igh}K$. Then, at $g=0$, we have
\begin{subequations}
\begin{align}
    \partial_g \widetilde{K}_0 &= -i dt \sum_{k,\lambda} q_{k,\lambda}L_{k,\lambda}^\dagger + O\left(dt^2\right) \\
    \partial_g \widetilde{K}_{k,\lambda} &= i q_{k,\lambda} \sqrt{dt} I + O\left( dt^{3/2} \right).
\end{align}
\end{subequations}
Defining $\alpha_{dt}$ and $\beta_{dt}$ as in Ref.~\cite{Zhou2018HNLS}, we thus have
\begin{subequations}
\begin{align}
    \alpha_{dt} &= \frac{m C}{\kappa} \, dt \, I + O\left(dt^2\right) \\
    \beta_{dt} &= O\left(dt^2\right).
\end{align}
\end{subequations}
where $C = \sum_\lambda |c_\lambda|^2$ is a constant with respect to $m$, $n$, and $\kappa$. We therefore have \cite{Sekatski:2017xdg}
\begin{align}
    F_Q \leq \frac{4mCT}{\kappa}
\end{align}
and thus
\begin{align}
    \frac{1}{\left( \Delta g\right)^2} = O\left( m\frac{T}{\kappa} \right),
\end{align}
which, combined with the previous subsection, proves that the $G \in \mathcal{S}$ scaling of Eq.~\eqref{eq:BosonicLinearHNLS} is an upper bound.

\subsection{A Maximal-Scaling Protocol when HNLS is Satisfied}

We now transition to showing that the bounds cannot be made parametrically stronger, as they can be achieved. We first present a construction for the $G \notin \mathcal{S}$ case, demonstrating that the bosonic linear version of the Heisenberg limit can always be reached in this case. This is simply a bosonic implementation of the construction in Ref.~\cite{Zhou2018HNLS}.

Since $G \propto x \notin \mathcal{S}$, we can choose some real unit vector $w = (w_x, w_p)$ with $w_x > 0$ that is orthogonal to $({\rm Re\,}u_\lambda, {\rm Re\,}v_\lambda)$ and $({\rm Im\,}u_\lambda, {\rm Im\,}v_\lambda)$ for all\footnote{In fact, for the simple case we present here, HNLS can only be satisfied if all noise acts along the same direction in the $x$-$p$ plane, or if there is no noise at all. We ignore this simplification here since it will not generally hold for higher-order signals and noises, whereas this approach generalizes straightforwardly.} $\lambda$. Now let
\begin{align}
    \alpha = \sqrt{\frac{n}{2m}}\left( w_x + i w_p \right).
\end{align}
With one noiseless ancillary qubit, we can then define the code states
\begin{subequations}
\begin{align}
    \ket{0_L} &= \ketC{\alpha}^{\otimes m}\ketQ{0} \\
    \ket{1_L} &= \ketC{-\alpha}^{\otimes m}\ketQ{1}.
\end{align}
\end{subequations}
These have expected total occupation $n/2$ (rather than $n$), and thus the error correction strategy below, applied sufficiently frequently, will ensure the system never has more than $n$ expected photons.

Let $P$ be the projector onto the code space. Then $PL_{k,\lambda}P = 0$ for all $k$ and $\lambda$, and moreover
\begin{align}
    P L_{k,\lambda}^\dagger L_{\ell,\kappa} P = C_{k,\lambda;\ell,\kappa} P,
\end{align}
where $C_{k,\lambda;\ell,\kappa}$ are some constants. Our code thus satisfies the Knill-Laflamme conditions \cite{Knill:1997mom} and therefore, at leading order in $dt$, frequent error correction fully eliminates our errors.

What remains is evolution under the error-corrected signal,
\begin{align}
    PGP = \sqrt{mn} \, w_x \left( \ket{0_L}\bra{0_L} - \ket{1_L}\bra{1_L} \right).
\end{align}
The initial state $(\ket{0_L} + \ket{1_L})/\sqrt{2}$ thus gives
\begin{align}
    F_Q = 4w_x^2 \cdot mnT^2 \sim mnT^2,
\end{align}
saturating the bound for $n \geq 1$. As we note in the main text, we do not prove achievability for $n \ll 1$, as $n$ becomes increasingly noise-dominated as $n \to 0$, leading to potentially different behavior.

\subsection{A Maximal-Scaling Protocol when HNLS is Not Satisfied}

Finally we turn to showing that the $G \in \mathcal{S}$ bound can be achieved. We again present an explicit construction saturating this bound, which in this case is simply a squeezed collective mode: place the same collective mode $b$ as above in a momentum-squeezed state with average photon count $n/2$ (leaving room for noise to add photons during the observation time); leave the $m-1$ orthogonal modes in vacuum. This gives an initial momentum variance
\begin{align}
    \Var(p_b)_{t=0} = \frac{1}{2}\left( \sqrt{\frac{n}{2}+1} - \sqrt{\frac{n}{2}} \right)^2 \sim \frac{1}{n+1}.
\end{align}

Since our noise direction was $k$-independent, its structure is carried through to the collective mode, and thus the Lindblad master equation gives
\begin{subequations}
\begin{align}
    \frac{d\expect{p_b}}{dt} &= -\kappa d \expect{p_b} - g\sqrt{m} \\
    \frac{d\,\Var(p_b)}{dt} &= -2\kappa d \, \Var(p_b) + \kappa D,
\end{align}
\end{subequations}
where we have defined
\begin{subequations}
\begin{align}
    d &= {\rm Im}\left( \sum\limits_\lambda u_\lambda^* v_\lambda \right) \\
    D &= \sum\limits_\lambda \left| u_\lambda \right|^2.
\end{align}
\end{subequations}
Note that $d$ is at most the number of noise types $|\{\lambda\}|$, and thus has an $(m,n,\kappa)$-independent upper bound. Although there is also noise on the $m-1$ other modes, we can frequently reset these to vacuum without affecting our signal mode such that they do not bring us above our $n$ photon ceiling.

We will integrate only for times $\kappa |d| T_{\rm int} \ll 1$, in which case
\begin{subequations}
\begin{align}
    \expect{p_b}(T_{\rm int}) &= -g\sqrt{m}T_{\rm int} + O\left(T_{\rm int}^2\right) \\
    \Var(p_b)(T_{\rm int}) &= \left(1-2\kappa d T_{\rm int}\right) \Var(p_b)_{t=0} + \kappa D T_{\rm int} + O\left(T_{\rm int}^2\right).
\end{align}
\end{subequations}
Then a momentum measurement gives (classical) Fisher information
\begin{align}
    F_C = \frac{ \left| \partial_g \expect{p_b}(T_{\rm int}) \right|^2}{\Var(p_b)(T_{\rm int})} \sim \frac{mT_{\rm int}^2}{\frac{1}{n+1} + \kappa T_{\rm int}}.
\end{align}

Now choose an integration time
\begin{align}
    T_{\rm int} \sim \min\left( T, \frac{c}{\kappa (n+1)} \right),
\end{align}
where $0<c<1$ is some constant with respect to $m$, $n$ and $\kappa$ that we do not compute exactly, repeating the measurement $T/T_{\rm int}$ times. The number of photons added in this integration time is $O(\kappa T_{\rm int} (n+1)) = O(c)$, and thus for sufficiently small $c$ this will keep us within our photon budget. Then the total QFI is
\begin{align}
    F_Q \sim m \min\left( (n+1)T^2, \frac{T}{\kappa} \right),
\end{align}
saturating our parametric bound.

\end{document}